\documentclass[12pt]{article}
\usepackage{axodraw4j}
\usepackage{color}
\usepackage{epsfig}
\usepackage{latexsym}
\usepackage{amssymb}
\usepackage{amsfonts}
\usepackage{pstricks}
\newcommand{\E}{\mbox{e}}
\newcommand{\be}{\begin{equation}}
\newcommand{\ee}{\end{equation}}
\newcommand{\bea}{\begin{eqnarray}}
\newcommand{\eea}{\end{eqnarray}}
\newcommand{\al}{\alpha}

\newcommand{\gm}{\gamma}
\newcommand{\Gm}{\Gamma}

\newcommand{\ep}{\varepsilon}

\newcommand{\dd}{\mbox{d}}

\newcommand{\nn}{\nonumber}
\newcommand{\D}{\mbox{d}}

\newcommand{\I}{i}
\newcommand{\insl}{\not\!\, \in}

\begin{document}
\parindent=1.5pc

\begin{titlepage}

\rightline{SFB/CPP-09-118, TTP09-39}
\bigskip
\begin{center}
{{\large\bf
{\tt FIESTA 2}: parallelizeable multiloop numerical calculations
} \\
\vglue 5pt
\vglue 1.0cm
{\large  A.V. Smirnov}\footnote{E-mail: asmirnov80@gmail.com}\\
\baselineskip=14pt
\vspace{2mm}
{\normalsize Scientific Research Computing Center, Moscow State
University, \\ 119992 Moscow, Russia
%\\
%Institut f\"{u}r Theoretische Teilchenphysik - Universit\"{a}t
%Karlsruhe
   }\\
\baselineskip=14pt
\vspace{2mm}
\baselineskip=14pt
\vspace{2mm}
{\large   V.A. Smirnov}\footnote{E-mail: smirnov@theory.sinp.msu.ru}\\
\baselineskip=14pt
\vspace{2mm}
{\normalsize
Skobeltsyn Institute of Nuclear Physics of Moscow State University, \\
119992 Moscow, Russia
%\\
%Institut f\"{u}r Theoretische Teilchenphysik - Universit\"{a}t
%Karlsruhe
%Moscow 119992, Russia
}\\
\baselineskip=14pt
\vspace{2mm}
{\large   M. Tentyukov}\footnote{E-mail: tentukov@particle.uni-karlsruhe.de}\\
\baselineskip=14pt
\vspace{2mm}
{\normalsize
Institut f\"ur Theoretische Teilchenphysik, Karlsruhe
  Institute of Technology, \\D-76128 Karlsruhe, Germany
}
\baselineskip=14pt
\vspace{2mm}
\vglue 0.8cm
{Abstract}}
\end{center}
\vglue 0.3cm
{\rightskip=3pc
 \leftskip=3pc
\noindent
The program {\tt FIESTA} has been completely rewritten. Now it can be used not only
as a tool to evaluate Feynman integrals numerically, but also to expand Feynman
integrals automatically in limits of momenta and masses with the use of sector
decompositions and Mellin--Barnes representations. Other important
improvements to the code are %the following:
complete parallelization (even to
multiple computers), high-precision arithmetics (allowing to calculate
integrals which were undoable before), new integrators and Speer sectors as a strategy,
the possibility to evaluate more general parametric integrals.
\vglue 0.8cm}
\end{titlepage}

\section{Introduction}

Originally sector decomposition in alpha (Feynman) parametric representations
of Feynman integrals was used as a tool for analyzing the convergence and
proving theorems on renormalization and asymptotic expansions of Feynman
integrals \cite{Hepp,theory,BM,BdCMPo,books1a}. At that time the so-called Hepp and Speer
sectors were introduced \cite{Hepp,Speer2}.
The goal of sector decomposition is to decompose the initial integration domain
into appropriate subdomains (sectors) and introduce, in each sector, new
variables in such a way that the integrand factorizes, i.e. becomes equal to a
monomial in new variables times a non-singular function.

Much later sector decomposition became a tool for evaluating Feynman
integrals. Initially it was introduced in \cite{BH} (see
Ref.~\cite{Heinrich} for a recent review) and was used to verify
several analytical results for multiloop Feynman integrals, including
three- and four-loop \cite{3box,3loop,4loop} results. Currently there
are also two public codes performing the sector decomposition ---
one code by Bogner and Weinzierl \cite{BognerWeinzierl} and the other by
two of the present authors \cite{FIESTA}. The latter one was named
FIESTA which stands for ``Feynman Integral Evaluation by a Sector
decomposiTion Approach''. During the last year {\tt FIESTA} was applied in
\cite{FIESTA-appl}.

Another problem which can be solved with the sector decomposition is the problem
of asymptotic expansion of Feynman integrals in momenta
and masses. One might apply the universal strategy of expansion by regions
\cite{BS,books1a} in this case, however it is not always simple to reveal regions relevant
for a given limit. It was recently suggested \cite{Pilipp} to combine, in such
situations, the method of Mellin--Barnes representation \cite{MB,books2} with
modern sector decompositions \cite{BH,BognerWeinzierl,FIESTA}. In fact, this
idea was exploited many years ago. For example, in \cite{BdCMPo} the asymptotic
expansion of Feynman integrals in various limits of momenta and masses was
studied using Mellin transform and Hepp \cite{Hepp} or Speer \cite{Speer2}
sectors. However,
%in situations where these sectors do not provide a
%factorization of the integrand, no results could be obtained. It turns our that
%the Hepp sectors as well as Speer sectors
%(successfully used in the past to prove various mathematical
%results on Feynman integrals \cite{Hepp,Speer2,Speer1,Po,Bergere,BM,books1})
Hepp and Speer sectors
are applicable only for Feynman integrals at Euclidean external momenta, i.e
with $\left(\sum q_i \right)^2<0$ for any partial sum. For the same reason,
these sectors are only applicable for studying asymptotic behavior of Feynman
integrals in limits typical of Euclidean space.

One more example of applying sector decompositions can be found in
Refs.~\cite{DP} where leading and subleading logarithms in asymptotic
expansions of Feynman integrals in the high-energy limit were studied. This approach
was successfully applied up to two-loop level\footnote{Sector
decompositions are successfully used not only for multiloop Feynman integrals but also for integrals contributing to real
radiation \cite{SDforRR}.}.
Let us also mention a recently proposed geometrical approach to sector decomposition
\cite{Kaneko:2009qx}. To implement the corresponding strategy on a computer
looks to be a rather nontrivial task. This is however very desirable because
the method promises to be the optimal one.

During the last year {\tt FIESTA} has been greatly improved in various aspects.
The code is capable of evaluating many classes of integrals that one would not
be able to evaluate with the original {\tt FIESTA 1}. Moreover the code can
now be applied to solve the problem of obtaining asymptotic expansions of
Feynman integrals in various limits of momenta and masses. Let us list the
new features in {\tt FIESTA 2}.

\begin{itemize}

\item
Asymptotic expansion of Feynman integrals. The current code can automatically
expand Feynman integrals in various limits of momenta and masses. We are aiming at a general strategy which could be used to automatically expand
a given Feynman integral in a given limit of momenta and masses. In the second
version of {\tt FIESTA} we used as in \cite{Pilipp}, the old idea of combining
Mellin--Barnes representation with practical sector decompositions.
Since the modern sector decompositions
\cite{BH,BognerWeinzierl,FIESTA} are applicable not only at Euclidean external
momenta but also when some kinematic invariants are zero and
some of them are of the same sign, this idea will work in this case.

\item
Parallelization. {\tt FIESTA 2} uses the features of {\tt Mathematica} 7.0
allowing to parallelize the {\tt Mathematica} part of the algorithm (however
the code can still be used with {\tt Mathematica} 6.0 in sequential mode). The
licensing policy of {\tt Mathematica} allows to launch up to 4 subkernels per
licensed kernel. And at clusters with license managers you should have normally
no problem with launching even more --- there are not many programs that might use
the subkernels.    \\
Moreover, the integration can be now paralleled to multiple computers via
TCP/IP. Examples show that the speed-up is about linear or even better (the
reason is that the main machine also performs some tasks related to the
database).

\item New methods to deal with numerical instability. We compare different ways
of treating singularities after the sector decomposition. There are basically two
ways: integration by parts and the Taylor expansion. We demonstrate that the second
one is preferable, however it results in the following problem: during the integration
the algorithm has to divide by sector variables in some powers resulting in huge numbers.
Although one might need a few valuable digits
for an integral, the intermediate calculations might be greatly improved if
they use high-precision arithmetics. This
improvement allows to proceed at the level which was
%integrate things that were
completely unreachable with the original {\tt FIESTA}.

\item Speer sectors. As explained in Ref.~\cite{SS09}, Speer sectors can be
used as an iterative strategy for sector decomposition. However they are
reproduced by Strategy~S. Still it is worth to mention that for complicated
examples Strategy~S might practically fail because of exponentially growing
time, and that is where Speer sectors should be used. We provide such an
example.

\item There are some other new features such as dealing with integrals
at the threshold, integrating the highest poles analytically, using
different integrators, the possibility to evaluate more general parametric integrals.
\end{itemize}

In the next sections we explain all the new features of the algorithm. They are
followed with installation instructions and the listing of all options of the
code. Finally we illustrate the usage of the new code by providing numerical
results for one of the most complicated massless four-loop propagator master
integrals \cite{BaCh}.

\section{Expanding Feynman integrals}

The $\alpha$-representation of Feynman integrals is a common technique
providing a possibility to express $d$-dimensional Feynman integrals
\cite{dimreg1,dimreg2} as classical integrals depending on the parameter of
dimensional regularization, $d=4-2\ep$ \cite{dimreg2,BM} Let us recall the
standard notation.

For a Feynman integral with standard propagators (of the type
$1/(m^2-k^2-i0)^{a_l}$) corresponding to a connected graph
$\Gm$, the alpha representation has the following form:
\bea
G_{\Gm}(q_1,\ldots,q_n;d;a_1\ldots,a_L)
%{\cal F}_{\Gm}(a_1\ldots,a_L;d)
& =& \frac{\I^{a+h(1-d/2)}  \pi^{h d/2}}{\prod_l \Gm(a_l)} \nn \\
&& \hspace*{-30mm}
\times \int_0^\infty\ldots\int_0^\infty \prod_l
\al_l^{a_l-1} {U}^{ -d/2}   \E^{\I {F}/{U}-\I\sum
m_l^2 \al_l} \dd\al_1 \ldots \dd\al_L \; ,
\label{alpha}
\eea
where  $L$ and $h$ is, respectively, the number of lines (edges)
and loops (independent circuits) of the graph, $n+1$ is the number
of external vertices, $a=\sum a_l$, and \bea {\cal U}& = &
\sum_{T\in T^1} \prod_{l\insl  T} \al_l
 \;,
\label{Dform}  \\
{F}&=& \sum_{T\in T^2} %\left[
\prod_{l\insl  T} \al_l
%\sum_i q_i
\left( q^T\right)^2
%\right]
 \; .
\label{Aform} \eea In (\ref{Dform}), the sum runs over trees
%{\em trees}\index{subject}{tree}
of the given graph,
%i.e. maximal connected subgraphs without loops,
and, in (\ref{Aform}), over {\em 2-trees}, i.e. maximal subgraphs
that do not involve loops and consist of two connectivity
components; $\pm q^T$ is the  sum of the external momenta that
flow into one of the connectivity components of the 2-tree $T$.
(It does not matter which component is taken because of the
conservation law for the external momenta.) The products of the
alpha parameters involved are taken over the lines that do not
belong to the given (2-)tree $T$. The functions $U$ and $F$ are homogeneous functions of the alpha parameters with the
homogeneity degrees $h$ and $h+1$, respectively.

Starting from (\ref{alpha}) one can introduce primary sectors
$\Delta_l$ by $\al_i\leq \al_l,\; i\neq l$. For example, in the
case of $l=L$, using the homogeneity properties of the functions
in the $\alpha$-representation and explicitly integrating over $\alpha_L$
one obtains the contribution of $\Delta_L$ as
\bea G^{(L)}
%{\cal F}_{\Gm}(q_1,\ldots,q_n;d)
& =&
 \frac{\left(\I\pi^{d/2} \right)^h
\Gm(a-h d/2)}{\prod_l\Gm(a_l)} \int_0^1 \ldots\int_0^1
\prod_l^{L-1} \al_l^{a_l-1}\,
\nn \\ && %\hspace*{-56mm}
\times \frac{ \hat{U}^{a-(h+1) d/2}} {\left[-\hat{F}
+\hat{U} \left(\sum_{l=1}^{L-1} m^2_l \prod_{l=l'}^{L-1}
\al_{l'}+m^2_L\right)\right]^{a-h d/2}} \dd \al_1 \ldots \dd
\al_{L-1} \;,
\label{primsec}
\eea
where
\be
\hat{U}={U}(\al_1,\ldots,\al_{L-1},1)\;, \;\;\;
\hat{F}={F}(\al_1,\ldots,\al_{L-1},1)\;.
\ee
Without
loss of generality let us consider only this primary sector.

%\newpage

We are dealing with a Feynman integral whose external momenta can
be fixed by some conditions, for example, put on a mass shell. Then some
terms contributing to $\hat{F}$ and massive terms can be
combined. Let us now suppose that we are studying a limit where
the kinematic invariants and the masses are decomposed into two
groups,
%we want to decompose the whole function depending on
%the external momenta and the masses into two pieces,
\be
\sum_{l=1}^{L-1} m^2_l \prod_{l=l'}^{L-1} \al_{l'}+m^2_L = W_1+W_2
\;, \ee and that those in the term $W_1$ are much smaller than
$W_2$. We introduce the parameter of expansion, $\lambda$ by
multiplying by it the terms of the first group.
%In particular, this can be studying a limit where one of the pieces is much larger than
%the other one.
Let us then separate the two group of terms by introducing a
onefold MB integral,
\be
\frac{\Gm(a-h d/2)}{(\lambda W_1+W_2)^{a-h d/2}}= \frac{1}{2\pi
i}\int_{-i \infty}^{+i \infty} \dd z\, \lambda^z \frac{\Gm(a-h
d/2+z) \Gm(-z)}{W_1^{-z} W_2^{a-h d/2+z}}  \; \, \label{MB1} \ee
so that we obtain \bea {\cal F}^{(L)} & =& \frac{\left(\I\pi^{d/2}
\right)^h}{\prod_l\Gm(a_l)} \frac{1}{2\pi i}\int_{-i \infty}^{+i
\infty} \dd z\, \Gm(a-h d/2+z) \Gm(-z) \lambda^z
\nn \\ && %\hspace*{-56mm}
\times\int_0^1 \ldots\int_0^1 \hat{\cal U}^{a-(h+1) d/2} \,
W_1^{z} \, W_2^{-a+h d/2-z}\, \prod_l^{L-1}\left( \al_l^{a_l-1}
\al_l \right)\,
%\dd \al_1 \ldots \dd \al_{L-1}
\;. \label{primsec1} \eea

The idea of using MB representation is to reduce the problem of
expansion to an analysis of poles in the variable $z$ of the
integrand. To pick up terms of expansion in the limit $\lambda\to
0$ one closes the integration contour to the right and takes
residues in $z$. (The residues are taken with the minus sign
according to the Cauchy theorem.) In addition to the poles of one
of the two explicitly present gamma functions  $\Gm(-z)$ at
$z=0,1,2,\ldots$, we have poles coming from the parametric
integration. In fact, we have to distinguish poles which are of
the same character as poles of gamma functions with $-z$
dependence.

Our algorithm which is implemented within of {\tt FIESTA} consists of the
following steps.

{\em Step 1.} The resolution of singularities of the integral over
$\al_1,\ldots,\al_{L-1}$ by a sector decomposition. Instead of the two
functions in the integrand of (\ref{primsec}), we have the three functions
$\hat{\cal U}$, $W_1$ and $W_2$ raised to certain powers depending not only on
$\ep$ but also (for $W_{1,2}$) on the MB integration variable $z$. As a result
of this procedure we obtain a sum of parametric integrals where all these three
functions are proper factorized, i.e. represented as powers of sector variables
times positive functions. Therefore each of resulting parametric integrals is
represented as an integral over $t_1,\ldots,t_{L-1}$ from $0$ to $1$ of $\prod
t_i^{r_i-1}$ times a product of positive functions raised to some powers. Here
exponents $r_i$ have the form $b_i \ep + c_i z +n_i$ where $b_i, c_i, n_i$ are
rational numbers.

%In the usual situation of evaluating Feynman integrals numerically
%by sector decompositions, one proceeds further with Taylor
%subtractions of sufficient order which provide convergence of
%integrals over $t_i$. The subtracted terms are integrated
%explicitly. We do not do this at this step yet, although this
%could be already done for the variables $t_i$ whose exponents are
%independent of $z$.

{\em Step 2.} Let us reveal singularities in $\ep$ generated by
the MB integration over $z$. The integral of $t_i^{b_i \ep + c_i z
+n_i-1}$ generates a $z$-dependence of the type $\Gm\left(b_i \ep
+ c_i z +n_i \right)$. We are concentrating on sector integrals
with $c_i<0$ because they are relevant to our limit.
%We are arriving at the problem of resolving singularities in $\ep$
%of the integral over $z$, where, in addition to the two gamma functions
%$\Gm(a-h d/2+z) \Gm(-z)$, we have $\Gm\left(b_i \ep + c_i z +n_i \right)$
%for each $t_i$-integration. Use a part of {\tt MBresolve.m} to do this.
%Then we obtain an integration over $z$ along a straight line with a fixed
%Re$z=x_0$ and a sum of terms which result from residues and involve
%no integration over $z$.
%
%Therefore we obtain two groups of terms: with and without integration over $z$.
%
%
%{\em Step 3.}

Using Taylor subtractions of sufficient order for the rest of the
integrand we decompose every integral over such $t_i$ into the
corresponding integral with a remainder and an integral of the
subtracted terms which is evaluated analytically.
%Observe that if $x_0$ is present in the exponent of a given $t_i$ then this
%integration does not generated poles in $\ep$ but the corresponding Taylor
%subtractions are needed to obtain integrals which can be treated numerically.
The remainder in such subtractions contributes to the
remainder of the whole expansion. When increasing the order of
expansion it tends to zero with a given power of
$\lambda$. The explicit integration of every Taylor
subtracted term provides a singular behavior in $z$  as a
rational function. We take residues in $z$ at these points,
similarly to the residues of the explicit $\Gm(-z)$.

{\em Step 3.} Every resulting residue is a sector integral where a
proper factorization due to the sector decomposition has been
achieved. It is treated numerically within {\tt FIESTA}.

The expansion mode of {\tt FIESTA} is able to handle complicated integrals. For
example, we studied the expansion of a massless on-shell box and a double box
when one of the Mandelstam variables $t$ is tending to zero.
The expansion was performed up to terms of order $t^0$ modulo logarithms and
the expansion
order of $\varepsilon$ was set to zero. The box was evaluated in
almost no time and the double box took less than half an hour.

\section{Parallelization}

A modern program should be able to take advantage of using multi-processor
computers or even multiple computers working together. This has been completely
implemented in {\tt FIESTA 2}. The parallelization now works both for the
{\tt Mathematica} part and for the integration.

\subsection{Parallelization in {\tt Mathematica}}

To enable parallelization in {\tt Mathematica} one has to use the 7.0 version.
Users of 6.0 should still be able to use {\tt FIESTA 2} in sequential
mode. The version 5.2 and lower version are not supported in {\tt FIESTA 2}.

To turn on the parallelization one has to set the {\tt NumberOfSubkernels} option to
something bigger than one. This will result in launching a number of subkernel
{\tt Mathematica} processes. The sector decomposition stage is performed by
sending jobs for primary sectors as different subkernel processes. After the sector
decomposition is over, one results in a number of sectors and all operations in
those might be performed absolutely independently, therefore, the problem
theoretically should be efficiently parallelized by {\tt Mathematica} 7.0 with
the basic {\tt Parallelize}\footnote{The {\tt Parallelize} function distributes
operations that have to be performed with all elements of a list over all
subkernels.} function.

However, in complicated cases it is impossible to keep
all the data in RAM, and already {\tt FIESTA 1} was able
to use {\tt QLink}\footnote{{\tt QLink} is a Mathlink based program by A.~Smirnov used to
store data from {\tt Mathematica} on disk. The original version used the {\tt
QDBM} database, now we have switched to the {\tt TokyoCabinet} database.}
to store its data on a hard disk. In {\tt FIESTA 2} only the main {\tt Mathematica} process accesses
the database and distributes evaluation tasks between subkernels. To do this
we had to use the queuing system of {\tt Mathematica} 7.0 directly, with the use of
functions that belong to private packages\footnote{The functions
{\tt ParallelSubmit}, {\tt WaitNext}, {\tt WaitAll}, {\tt Parallel`Developer`QueueRun[]}
as well as using the structure of evaluation object directly and comparing
a part of those with {\tt Parallel`Developer`finished}.}.

The queueing system and the disk access might be considered as a bottleneck for
parallelizing; for tasks that might be performed in less than a few minutes you
might even result in a slowdown by turning the parallel mode on. Still for
complicated examples it turns out that the parallelization of this type is
really efficient. Here is a table demonstrating the comparison for
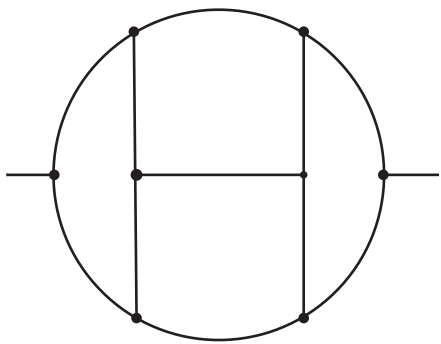
\begin{figure}
\begin{center}
\fcolorbox{white}{white}{
  \begin{picture}(165,126) (43,-15)
    \SetWidth{1.0}
    \SetColor{Black}
    \Arc(124,48)(62.29,312,672)
    \Line(156,101)(156,-6)
    \Line(92,101)(93,-5)
    \Line(44,48)(62,48)
    \Vertex(93,-6){2}
    \Vertex(62,48){2}
    \Vertex(92,102){2}
    \Vertex(186,48){2}
    \Vertex(156,-6){2}
    \Vertex(156,102){2}
    \Line(186,48)(207,48)
    \Line(93,48)(156,48)
    \Vertex(93,48){2.236}
    \Vertex(156,48){1.414}
  \end{picture}
}
\end{center}
\caption[]{One of the most complicated four-loop propagator diagrams.}
\label{m61}
\end{figure}
the massless propagator Feynman diagram with 11 lines shown in Fig.~\ref{m61}:

\noindent
\begin{tabular}{|l|c|c|c|c|c|}
  \hline
  Stage                            & time for 1-kernel &  time for 8-kernel  &
speedup\\
  \hline
  Variable substitution                         &  1025 &   368  & 2.79 \\
  Decomposing $\ep$-independent term            &   386 &   78  & 4.95 \\
  Pole resolution                               &   792 &   126  & 6.29 \\
  Expression construction                       &   540 &   115  & 4.70 \\
  Epsilon expansion                             &   531 &   112  & 4.74 \\
  Expansion, string preparation                 &       &        &      \\
                              for order -1      &  1118 &   152  & 7.36 \\
  Expansion, string preparation                 &       &        &      \\
                             for order 0        & 46246 & 6110   & 7.57 \\
\hline
\end{tabular}

You can see that for most time-consuming operations we are approaching the
unreachable factor of 8.

The integration can also be performed with {\tt Mathematica} and then it can be also
parallelized. However it is not clear how to measure the speedup and other factors.
The reason is that one has no full control of evaluation precision and the
number of sample points during the integration within {\tt Mathematica}. If one
wishes to integrate with {\tt Mathematica}, sometimes it makes reason to set
the {\tt MixSectors} options to something big or even infinity in order that
{\tt Mathematica} would perform integrations not within sectors, but with mixing
expressions beforehand.

\subsection{Parallelizing all stages of the algorithm}

Although the integration might be performed with {\tt Mathematica}, we strongly
suggest to use the external integration in C. It has many benefits that we will
show up below. The C integration was already
parallelized in {\tt FIESTA 1}, and the mechanism is rather
straightforward --- {\tt Mathematica} launches a number of CIntegrate
processes and communicates with them via the Mathlink protocol
distributing the individual integration jobs.

Altogether both Mathematica part and C integration demonstrate a good
scalability on modern multicore computers. Here are timings (in
seconds) for the same diagram of Fig.~\ref{m61}.

\noindent
\begin{tabular}{|c|c|c|c|c|}
\hline
              &                    & numerical         & numerical        & \\
NumberOfLinks & NumberOfSubkernels & $\varepsilon^{-1}$ & $\varepsilon^{0}$ & total \\
\hline
1&              0&                     5773.88&          95406.11&        160738.0\\
2&              2&                     2909.61&          47963.15&        81175.6\\
3&              3&                     1955.95&          32259.22&        54444.1\\
4&              4&                     1471.02&          24759.72&        41578.9\\
5&              5&                     1178.59&          20526.16&        34614.0\\
6&              6&                     1004.62&          17659.31&        29391.1\\
7&              7&                     855.03 &          15585.42&        26028.6\\
8&              8&                     765.57 &          14087.94&        23910.7\\
\hline
\end{tabular}

\noindent The results were obtained on a Xeon E5472 3.00GHz 8-core
computer. The first line (NumberOfLinks = 1, NumberOfSubkernels = 0)
is a ``reference point'' since all the job is performed by only one
worker. The next line (NumberOfLinks = 2, NumberOfSubkernels = 2) is
the next step in parallelization since the job is done by two workers
simultaneously. The last line corresponds to the case when all
available CPU cores work in parallel. The third and the forth
columns contain times for numerical C integration of
$\varepsilon^{-1}$ and $\varepsilon^{0}$ terms, correspondingly. The
last column shows the total time spent by FIESTA to calculate the
integral up to a finite part (in $\ep$).

We also made the same benchmark on an AMD Opteron Processor
2439 2.8 GHz 12-core computer. Here are results:

\noindent
\begin{tabular}{|c|c|c|c|c|}
\hline
              &                    & numerical         & numerical        & \\
NumberOfLinks & NumberOfSubkernels & $\varepsilon^{-1}$ & $\varepsilon^{0}$ & total \\
\hline
1&             0&                    7791.40&         103471.28&      174601.0\\
2&             2&                    3969.19&         51972.78 &      88158.4\\
4&             4&                    1992.81&         26390.38 &      44664.9\\
5&             5&                    1628.65&         21261.42 &      36698.4\\
6&             6&                    1358.80&         17817.81 &      30591.3\\
7&             7&                    1162.79&         15332.68 &      26629.6\\
8&             8&                    1037.65&         13599.75 &      23784.6\\
9&             9&                    928.76 &         12314.39 &      21615.8\\
10&            10&                   831.43 &         11067.77 &      19680.6\\
11&            11&                   771.11 &         10229.74 &      18269.2\\
12&            12&                   724.57 &         9755.14  &      16788.5\\
\hline
\end{tabular}

\begin{figure}[ht]
  \centering
  \includegraphics[width=\textwidth]{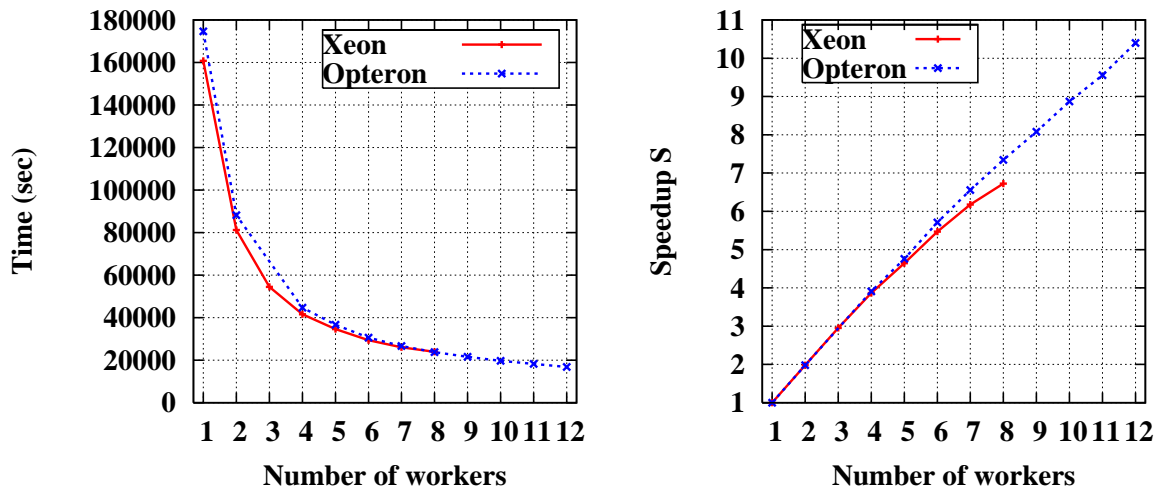}
  \caption{\label{speedup} Absolute time and speedup for Xeon and Opteron computers.}
\end{figure}

Total times are plotted in Fig.~\ref{speedup} together with a speedup
$$S(p)=\frac{\mbox{T(NumberOfLinks=1,NumberOfSubkernels=0)}}{\mbox{T(NumberOfLinks=p,NumberOfSubkernels=p)}}$$.

As we can see, FIESTA scales on a Xeon computer essentially
worse than on an Opteron one. Note that in both cases we have tested
absolutely the same algorithm and software. The main hardware
difference is a communication medium between processors and memory:
Xeon has a shared Front-Side bus while Opteron uses point-to-point
``hypertransport'' mechanism.

Numerically the communication overhead can be very roughly estimated
using the following formal model (see, e.g., \cite{Horoi}).

Let $\alpha$ be an essentially non-parallelizeable part of the
job. The time spent by one worker is  $T_1=\alpha T_1 + (1-\alpha)
  T_1$, and $p$ workers time is $T_p = \alpha T_1 +
  \frac{(1-\alpha)T_1}{p}+t_d$,
where $t_d$ is the communication overhead.
The speedup  $S(p)=\frac{T_1}{T_p}=
\frac{T_1}
  {
   \left(
      \frac{
         1-\alpha
      }{
         p
      }
      +\alpha
   \right)
   T_1+t_d
  }
$,
$t_d=0$ corresponds to the Amdahl's law \cite{Amdahl}:
$S(p)= \left( \alpha +\frac{1-\alpha}{p} \right)^{-1}$.

Let us assume that
the Opteron computer has a ``good'' communication medium,\
$t_d=\mbox{const}=b$, while the Xeon has a ``bad'' one,
$t_d=b_1+b_2\times p$. We would like to fit numerically the Amdahl
constant from the Opteron data and use it to estimate $b_1$ and $b_2$
for the Xeon data.
If $t_d=\mbox{const}=b$ then it can be absorbed by the Amdahl
constant $\alpha$. Indeed, let $\beta=\alpha+\frac{b}{T_1}$, then
$S(p)=\frac{1}{
\frac{1-\beta}{p}+\beta+\frac{b}{p T_1}
}$. Assuming $b<<T_1$, the term $\frac{b}{p T_1}$ is negligible and
will become zero by any good fit algorithm.

Fitting the Opteron data, we obtain $\alpha = 0.0016$ (only 0.16 \% is
non-parallelizeable, including communication overhead) and then for
the Xeon we get $b_2=428.305$ seconds per a CPU core, and $b_1$ is
negligible. This means, running this FIESTA job with 8 CPU cores in
parallel, the Xeon computer spends about 14 \% of total time for
communications.

\begin{figure}[ht]
  \centering
  \includegraphics[width=.5\textwidth]{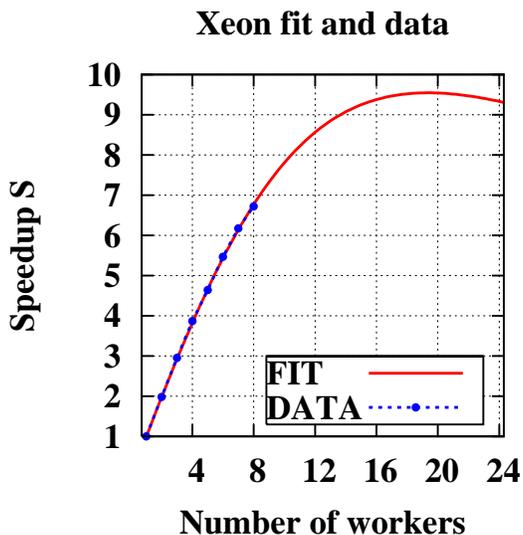}
  \caption{\label{AmdahlFit} Data and extrapolated result for the Xeon fit.}
\end{figure}

The fitted speedup function for the Xeon computer has a maximum at 19
workers, and then it goes down.  In Fig.~\ref{AmdahlFit} the fitted
curve is extrapolated to the region beyond the theoretical maximum for
this model which is only 9.54.

{\tt FIESTA 1} is able to parallelize the C integration only on
multicore computers. {\tt FIESTA 2} goes further, in the sense that one might set up a
network of computers performing the integration. The communication is again
performed via the Mathlink protocol since it can use not only the shared memory but
also TCP/IP for data exchange.

One should not worry about the network traffic. Most of the time the processes
work independently only exchanging keep-alive packets between each other.
Moreover, the setup is relatively safe: if the network goes down, the main
process keeps distributing tasks on a local machine and at the end
recalculates the parts that were sent to remote machines and returned no
answer. If a single integration process goes down it also does not spoil the
calculation: {\tt FIESTA} simply removes that particular process from the list
of workers and resubmits its task elsewhere.

The tests show that moving, for example, from one 8-kernel machine to two
8-kernel machines gives more than a double speedup. The reason is that the main
machine also works with a hard drive, while the other one is free from that.

To set up a calculation on multiple computers one has to launch slave tasks
first. To do this one has to run {\tt CIntegrate -slave FILENAME} on slave
machines. (Usually one launches the number of slave processes equal to the
number of kernels on the slave machine.) The {\tt FILENAME} is the name of a
file where the slave writes down some information about the IP-address of the slave
machine and the two ports it is listening to. Unfortunately it is impossible to
control the exact port number, therefore one might encounter problems if
firewalls are turned on in the cluster. The file should be accessible from all
the machines you wish to use in the parallel evaluation. Afterwards one can load
{\tt FIESTA} on the main machine, set {\tt RemoteLinksFile} variable equal to
the path to that file and launch the task.

\section{Numerical instability}
After the resolution of singularities, one often encounters functions like
\begin{eqnarray}
\int_{0}^1 \ldots\int_{0}^1 \dd x_i\ldots \dd x_{n}\left(\prod_{j=1}^n x_j^{a_j-1+b_j \ep}\right) Z,
\label{NoSing}
\end{eqnarray}
where $Z$ has no singularities. Such a function still might have
singularities because of non-positive $a_j$.

Let us assume that $a_i\leq0$ and treat the integrand as a function
$x^{a_i-1+b_i \ep} Y(a_i)$
with coefficients being polynomials in other variables. For readability we will
omit the index $i$:
\be
\label{negPowrs}
x^{a-1+b \ep} Y(x)
\ee
To integrate such a function, the original {\tt FIESTA} replaces
(\ref{negPowrs}) by
\begin{eqnarray}
Y(0)x^{a-1+b \ep}+Y'(0)x^{a+b
  \ep}+\ldots+\frac{1}{a!}Y^{(-a)}(0)x^{-1+b \ep}+
\nonumber\\\nonumber
x^{a-1+b \ep} \left( Y(x)-Y(0)-Y'(0)x-\ldots-\frac{1}{a!}Y^{(-a)}(0)x^{-a}\right)
\end{eqnarray}
The items in the first line of the expression can be
integrated analytically over $x$ leaving us with one integration less. As for the
remainder, it is known that it has no singularities at $x=0$ hence it
can be expanded in $\ep$ and integrated. However we result in
integrating an expression that is a sum of potentially huge terms for
small values of integration variables which sometime result in a
nonsense when evaluated by a computer.
As a workaround {\tt FIESTA 1} had the so-called IfCut's ---
replacements of the integrand with its first terms of Taylor series
for small values of integration variables. However, such an approach
significantly slows the evaluation down and, more importantly, results in
uncontrollable error estimates. Moreover, for some complicated
situations problems start to appear even for really big values of
integration variables, in pathological cases even setting the cut to $0.5$
(that is actually integrating something really different from the original function)
makes this approach fail!

\subsection{Resolving numerical instability by integration by parts}

An alternative way to deal with expressions (\ref{negPowrs}) is to use
several times the integration by parts (IBP) formula in order to make the
power of $x$ positive.  However this method results in taking an extra
derivative, and therefore all expressions become more complicated.

{\tt FIESTA 2} has an option {\tt ResolutionMode}. The default value of
this option is {\tt Taylor} which reproduces the {\tt FIESTA 1} Taylor
expansion. If one sets {\tt ResolutionMode} to {\tt IBP0} then
{\tt FIESTA 2} uses IBP with the representation
$x^{a-1+b \ep} = -\frac{\dd}{{\dd}x}\int_x^1 t^{a-1+b \ep} \dd t$.
Since $\int_1^1 t^{a-1+b \ep} = 0$, the
surface term is equal to $Y(0)$. The third
possibility for {\tt ResolutionMode} is {\tt IBP1}, in this case
$x^{a-1+b \ep} =\frac{\dd}{{\dd}x}\int_0^x t^{a-1+b \ep} \dd t$
and the surface term is $Y(1)$.

Both of these resolution methods solve the problem of numerical
instability but the efficiency might become very poor.

\begin{figure}[ht]
\begin{center}
\epsfxsize=.8\linewidth \epsfbox{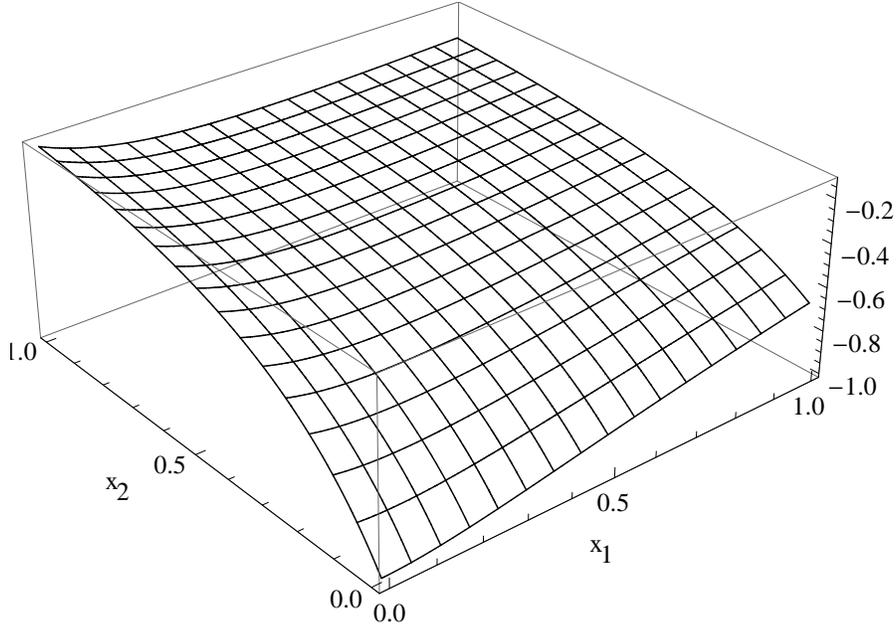}
\end{center}
\caption{
\label{TaylorPlt}
   3D-plot of integrand obtained with {\tt ResolutionMode} = {\tt     Taylor}, (\ref{Tailorin})
}
\end{figure}

Let us consider the following simple example\footnote{In fact, this is the propagator integral
with the one line index equal to 2 but we would like to consider it as
a vertex with one external momentum equal to zero since this example
is rather expressive for our purposes.}: the massless triangle
with one zero external momentum so there is only one external momentum
$p$, $p^2=-1$. Up to the constant part, the algorithm produces only six
integrations. The most complicated integrand is the ``number six'' one.
For {\tt ResolutionMode} = {\tt IBP1} it looks like follows:
\begin{eqnarray}
\nonumber&&\frac{\log(1 + x_1)}{(1 + x_1)(2 + x_1)}
            -\frac{2\log(2+x_1)}{(1+x_1)(2+x_1)}\\
\label{IBP1in}&& +\frac{2}{(1+x_2)(1+x_1+x_2)^2}
            +\frac{\log(x_1)}{(1 + x_2)(1 + x_1 +  x_2)^2}\\
\nonumber&& +\frac{\log(1+x_2)}{(1 + x_2)(1 + x_1 + x_2)^2}
            -\frac{2\log(1+x_1+ x_2)}{(1+x_2)(1 + x_1 + x_2)^2}
\end{eqnarray}
and  for {\tt ResolutionMode} = {\tt Taylor} it has the following form:
\begin{equation}
\label{Tailorin}
-\frac{\log(1+x_1)}{(1 + x_1)^2}-\frac{1}{x_1(1+x_2)^2}+\frac{1}{x_1(1+x_2)(1+x_1+x_2)}
\end{equation}
At small $x_1$, the last two terms in (\ref{Tailorin}) form a difference of
two large numbers. Moreover, both these terms diverge at $x_1\to 0$
but their difference at $x_1=0$ is a smooth function. Indeed,
contracting GCD (greatest common divisor) we obtain
\begin{equation}
\label{GCDoff}
-\frac{1}{x_1(1+x_2)^2}+\frac{1}{x_1(1+x_2)(1+x_1+x_2)}=
-\frac{1}{(1 + x_2)^2(1 + x_1 + x_2)}.
\end{equation}
Since there are only two integration variables, the function may be
visualized as a 3D-plot, see Fig.~\ref{TaylorPlt}.
The function is very smooth and well suited for integration.

\begin{figure}[ht]
\begin{center}
\epsfxsize=.8\linewidth \epsfbox{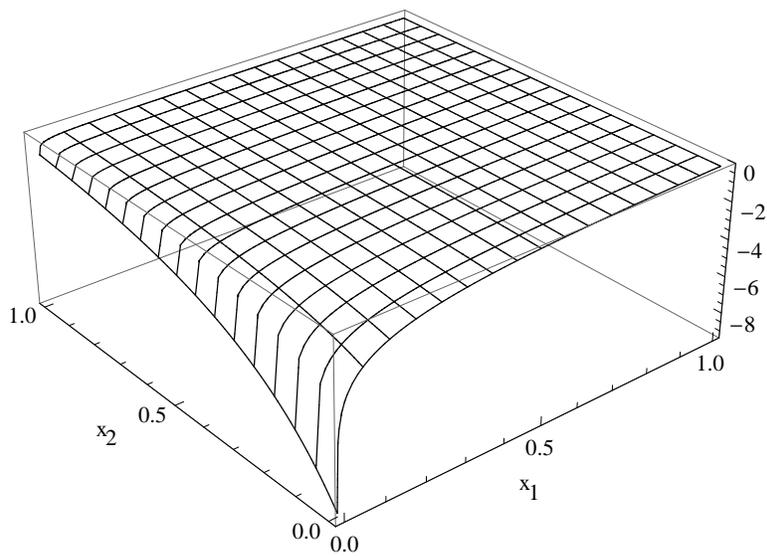}
\end{center}
\caption{
\label{IBP1Plt}
   3D-plot of integrand obtained with {\tt ResolutionMode} = {\tt     IBP1}, (\ref{IBP1in})
}
\end{figure}

The function(\ref{IBP1in}) has no non-integrable terms by
construction. Nevertheless, it contains $\log(x_1)$ and the shape of this
function is much more complicated for integration, see
Fig.~\ref{IBP1Plt}.

This is the common situation: the IBP methods produce much more terms, and
an integrand behaves worse than that produced by the Taylor method.
In fact, both IBP methods fail for complicated cases.

\subsection{High-precision arithmetics}
\label{highPrecision}

For technical reasons, one cannot make operations like (\ref{GCDoff})
automatically. First of all, a GCD contraction is a very time-consuming operation.
But more importantly, in a general case the Taylor approach results in expressions
containing some logarithms after the $\varepsilon$ expansion, hence it is impossible to
simplify the expression so that the large terms are canceled out.
For example, some terms may contain factors like $\log(1+x)/x$.

Let us consider what happens during the integration more precisely.

For instance, let $x_1$ be $10^{-20}$. Then the last two terms in (\ref{Tailorin}) are
of order of $10^{20}$ but their difference is about $10^{-1}$. This
means that all 20 leading decimal digits in fractions compensate each
other and the result is obtained only from the difference in the
fraction digits starting from 21. But in double precision IEEE
arithmetics a fraction has only 14-15 reliable digits.

Provided that the fraction has 26 reliable decimal digits, the result
contains 5 correct digits which is usually enough. So the natural idea
appears, namely, to use {\tt ResolutionMode} = {\tt Taylor} together
with a multiple precision floating-point arithmetics.

We have tested many different multiple precision floating-point
libraries. The best one in our case appears to be the GNU mpfr
library (http://www.mpfr.org).

One has to take into account that high-precision arithmetics libraries
are much slower than the native hardware floating-point arithmetics,
even for the same precision. Moreover, the larger precision is required, the
slower calculations are. So there are two problems that arise:
\begin{enumerate}
\item Which points must be evaluated with the high-precision arithmetics
and which ones might still remain in the native arithmetics?
\item Which precision must be set for the high-precision arithmetics?
\end{enumerate}

In order to answer the first question let us try to find the worst case.

It is known by construction that all problems come from the
negative powers of integration variables. Let us count maximal
negative powers $max_i$ for each variable $x_i$ and construct the artificial
auxiliary monomial
\begin{equation}
\label{testmonom}
\prod_{i=1}^n x_i^{max_i}
\end{equation}
where non-positive $max_i$ is the maximal negative power for $x_i$, or
zero, if there are no negative powers for $x_i$.
This monomial is the worst case in the sense that there are definitely
no more dangerous terms.

Before evaluating the integrand at a
concrete point $\vec x$ we calculate the monomial (\ref{testmonom}) at
this point. If the result is larger than some threshold we might be in a
trouble. In the native double precision IEEE arithmetics the fraction has
14-15 reliable decimal digits, so the default value for this threshold
is $10^9$ in order to have 5-6 reliable decimal digits in the result.
This value can be changed by means of the option {\tt MPThreshhold}. If
(\ref{testmonom}) at the point $\vec x$ is greater than the threshold
value, the integrand at the point $\vec x$ is evaluated using
the high-precision arithmetics.

Remember \cite{FIESTA} that the C-part of {\tt FIESTA}
is an interpreter of a string representing an expression to be
integrated. First, it compiles the integrand in some internal
representation and then it integrates the function evaluating it many times.
Precision for the high-precision arithmetics is fixed at the compile
time after the monomial (\ref{testmonom}) is built. To do this, we again
try to estimate the worst case.

We know that the function is smooth enough by construction, so a small shift
in $\vec x$ should not change the result much.

Assume that the worst case would be when all $x_i$ are equal to
some small value (0.001 by default, the user may
change this value by the option {\tt SmallX}). The exponent of (\ref{testmonom}) at
this point roughly gives us the number of digits in the fraction we might lose.

Let $M$ be the value of (\ref{testmonom}) at {\tt SmallX}.  Since
all numbers in a computer are of a binary format, the precision we need
also is in bits (the IEEE double precision fraction has 53 bit).
Calculating a binary logarithm of $M$ we get a number of bits in
fractions we might lose, so then we have to add some bits for digits we
would like to have in the fraction. The exact formula used by {\tt FIESTA} in
order to get the precision $p$ in bits is the following:
\begin{equation}
\label{precision}
   p=\log_2(M)+s
\end{equation}
where $s$ is the number of reliable bits we would like to have in the
fraction. The default value for $s$ is 38, it may be changed setting
the option {\tt PrecisionShift}.

Let us note that the user has the possibility to force {\tt FIESTA} to use some fixed
precision specifying the option {\tt DefaultPrecision}, e.g. setting
{\tt DefaultPrecision} = 1024 forces the multiple precision to be 1024
independently on the result of evaluation (\ref{testmonom}) at {\tt SmallX}.

As it was described above, the monomial (\ref{testmonom}) is calculated
at each point $\vec x$ before evaluating the integrand. It might
happen that the value of this monomial will be even more than $M$,
which means the precision (\ref{precision}) might not be enough. In
this case the algorithm increases all $x_i$ which are smaller than
{\tt SmallX} so that the value of (\ref{testmonom}) becomes not more
than $M$. For this step, the user is able to specify the value which
differs from $M$ by means of the option {\tt MPMin}. The reason to have such an
option is the following: for an expression like (\ref{negPowrs}) we
have essentially three different regions of our integration domain:
\begin{enumerate}
\item Native arithmetics.
\item High-precision arithmetics.
\item High-precision is not enough.
\end{enumerate}
Sometimes it is useful to have the possibility to set the last region manually.

Let us summarize the high-precision arithmetics options here. Note
that
all these options are relevant only if the option {\tt ResolutionMode}
is set to {\tt Taylor}.
\begin{itemize}
\item {\tt MPThreshhold} is the value of (\ref{testmonom}) when the
  algorithm switches to high-precision arithmetics, default is $10^9$.
  This is essentially a big value so if the user specifies
  $\mbox{\tt MPThreshhold} < 1$ then the algorithm uses the inverse value,
  $1/\mbox{\tt MPThreshhold}$. The bigger this value the better
  performance.
\item {\tt SmallX}
  is the value of $x$, $x_1=x_2=...=x_n=\mbox{\tt SmallX}$, where
  the value of (\ref{testmonom}) is evaluated in
  order to get the precision by the formula
  (\ref{precision}), default is 0.001. This is essentially a small
  value so if the user
  specifies $\mbox{\tt SmallX} > 1$ then the algorithm uses the
  inverse value, $1/\mbox{\tt SmallX}$. The bigger this value the
  better performance.
\item {\tt PrecisionShift} is the number of reliable bits in
  intermediate calculation, the value of $s$ in (\ref{precision}), the
  default value is 38 which roughly corresponds to 11 decimal
  digits. The smaller this value the
  better performance.
\item {\tt DefaultPrecision}. If the user specifies this option then
  the algorithm uses this precision instead of (\ref{precision}).
  The minimal value for {\tt DefaultPrecision} depends on the MPFR
  implementation, for mpfr-2.4.0 it is 2. The maximal value for {\tt
    DefaultPrecision} is 2147483647.
  Setting {\tt DefaultPrecision}=0 switches back to the default behaviour.
\item {\tt MPMin}. If the user specifies this option then
  the algorithm uses it instead of (\ref{testmonom}) at {\tt SmallX}
  in order to check whether the high precision is enough or the
  current $\vec x$ should be cut of. This is essentially a big
  value so if the user specifies  $\mbox{\tt MPMin} < 1$ then the
  algorithm uses the inverse value,
  $1/\mbox{\tt MPMin}$. Setting $\mbox{\tt MPMin} < 0$ switches back
  to the default behaviour. Changing this value does not influence the
  performance, the only reasons to set this option are debugging and
  profiling the code.
\end{itemize}

Note that all these options should be used with care. Playing with them,
it is very easy to get a wrong result, e.g. increasing the value {\tt
  MPThreshhold} the user may get better and better performance but
suddenly the result might become completely wrong.

%\subsection{Examples}

\section{Speer sectors}

\label{SectionSpeer}

As it has been explained in \cite{SS09}, Speer sectors can be also used as an
iterative strategy for sector decompositions. However to use it one has to
provide not only the propagators but also the diagram structure. Moreover, the
Speer sectors are only applicable at all Euclidean external momenta. We have
shown in that paper that in those cases our strategy S results in absolutely
the same set of sectors. One might ask a reasonable question: why would one
wish to use Speer sectors as a sector decomposition strategy at all? The answer
is rather simple: since the Speer sectors strategy knows more information it
can use it and work more efficiently, therefore performing the sector
decomposition faster. For simple cases there is no reason to use them --- one
would spend more time providing the diagram structure correctly that one would
win for the evaluation, but in some situations it might become crucial. Even
more, the time and RAM required to make a sector decomposition step for
strategy~S grows exponentially with the complexity of the problem. In some
cases the implementation of strategy S makes a failback to the non-efficient but
straight-forward strategy A\footnote{It is hardcoded that a sector decomposition
time limit for strategy S is 1000 seconds. We decided to do it in such
a way to
prevent the freezing of the program. However, theoretically strategy S is guaranteed
 to succeed.}. Therefore in some complicated examples the implementation of
strategy S might result in more sectors than an optimal strategy would
produce, or, strategy S might even fail. To demonstrate this we provide
examples of vacuum diagrams with equal masses on all lines, one vertex in the
center and $N$ vertices around.

\begin{center}
\fcolorbox{white}{white}{
  \begin{picture}(434,82) (-1,-31)
    \SetWidth{1.0}
    \SetColor{Black}
    \Line(0,-30)(48,50)
    \Line(48,50)(96,-30)
    \Line(96,-30)(0,-30)
    \Line(0,-30)(48,2)
    \Line(48,2)(96,-30)
    \Line(48,2)(48,50)
    \Line(128,50)(128,-30)
    \Line(128,-30)(208,-30)
    \Line(208,-30)(208,50)
    \Line(208,50)(128,50)
    \Line(128,50)(208,-30)
    \Line(208,50)(128,-30)
    \Line(256,-30)(304,-30)
    \Line(304,-30)(320,18)
    \Line(256,-30)(240,18)
    \Line(240,18)(280,50)
    \Line(320,18)(280,50)
    \Line(368,-30)(416,-30)
    \Line(368,-30)(352,10)
    \Line(352,10)(368,50)
    \Line(368,50)(416,50)
    \Line(416,50)(432,10)
    \Line(416,-30)(432,10)
    \Vertex(48,2){2}
    \Vertex(168,10){2}
    \Vertex(392,10){2}
    \Vertex(280,5){2}
    \Line(368,50)(416,-30)
    \Line(368,-30)(416,50)
    \Line(352,10)(432,10)
    \Line(256,-30)(280,5)
    \Line(240,18)(280,5)
    \Line(280,50)(280,5)
    \Line(320,18)(280,5)
    \Line(304,-30)(280,5)
  \end{picture}
}
\end{center}

The following table compares the time and the number of sectors produced by
strategies~S and the Speer sectors strategy SS. Please, keep in mind that the
diagrams have lots of internal symmetries, so actually only two primary sectors
should be considered for each of those. To take this into consideration we used
the {\tt PrimarySectorCoefficients} option. For example, for the most
complicated diagram, the hexagon, this option was set to
{\tt\{6,0,0,0,0,0,6,0,0,0,0,0\}} specifying that only the first and the seventh
primary sectors should be considered and the results multiplied by $6$. The
results are compared in the following table:

\begin{tabular}{|c|c|c|c|c|}
  \hline
  $N$   & Strategy S - time &   Strategy SS - time &  Number of sectors \\
  3 &   0.8 &    0.6 &    32 \\
  4 &    39 &     27 &   261 \\
  5 &  1859 &   1079 &  2574 \\
  6 &     F &  40570 & 29450 \\
  \hline
\end{tabular}

At level $6$ the strategy $S$ could not produce a result because of memory
overflow on a $2GB$ machine. (The sector decomposition stage does not use the
hard drive and normally requires almost no RAM.)

Now let us provide instructions how to use strategy SS. As it has been said
earlier, it is insufficient to set {\tt STRATEGY} to {\tt STRATEGY\_SS} since
the strategy has to use the information on the diagram. The evaluation is
started not with

{\tt SDEvaluate[\{U,F,l\},indices,order]},

but with

{\tt SDEvaluateG[graph\_information,\{U,F,l\},indices,order]}.

The graph information should be of the form $\{lines,external\_vertices\}$,
where $lines$ is a list of pairs of vertices connected by this line. The
vertices should be numbered from $1$ without skipping numbers. It is also very
important to have the order of lines coincide with the order of propagators in
the $UF$ input. For example, for the hexagon the input is:

\begin{eqnarray}\nonumber
&&SDEvaluateG[\{\{\{1, 2\}, \{2, 3\}, \{3, 4\}, \{4, 5\}, \{5, 6\}, \{6, 1\},
\\\nonumber
&&\{7, 2\}, \{7, 3\}, \{7, 4\}, \{7, 5\}, \{7, 6\}, \{7, 1\}\}, \{\}\},
\\\nonumber
&&UF[\{k1, k2, k3, k4, k5, k6\},
\\\nonumber
&&\{-k1^2 + 1, -k2^2 + 1, -k3^2 + 1, -k4^2 + 1, -k5^2 + 1, -k6^2 + 1,
\\\nonumber
&&-(k1 - k2)^2 + 1, -(k2 - k3)^2 + 1, -(k3 - k4)^2 + 1,
\\\nonumber
&&-(k4 - k5)^2 + 1, -(k5 - k6)^2 + 1, -(k6 - k1)^2 + 1\},
 \{\}],
\\\nonumber
&& \{1, 1, 1, 1, 1, 1, 1, 1, 1, 1, 1, 1\}, -1]
\end{eqnarray}

The diagram with all vertices numbered and all momenta labeled follows.
The notation coincides with the notation of the input for {\tt FIESTA}.

\begin{center}
\fcolorbox{white}{white}{
  \begin{picture}(292,224) (-1,-11)
    \SetWidth{1.0}
    \SetColor{Black}
    \Line[arrow,arrowpos=0.5,arrowlength=5,arrowwidth=2,arrowinset=0.2](32,96)(80,192)
    \Line[arrow,arrowpos=0.5,arrowlength=5,arrowwidth=2,arrowinset=0.2](80,192)(208,192)
    \Line[arrow,arrowpos=0.5,arrowlength=5,arrowwidth=2,arrowinset=0.2](208,192)(256,96)
    \Line[arrow,arrowpos=0.5,arrowlength=5,arrowwidth=2,arrowinset=0.2](256,96)(208,0)
    \Line[arrow,arrowpos=0.5,arrowlength=5,arrowwidth=2,arrowinset=0.2](208,0)(80,0)
    \Line[arrow,arrowpos=0.5,arrowlength=5,arrowwidth=2,arrowinset=0.2](80,0)(32,96)
    \Vertex(144,96){3}
    \Vertex(32,96){3}
    \Vertex(80,192){3}
    \Vertex(208,192){3}
    \Text(32,112)[r]{\Large{\Red{$1$}}}
    \Text(64,192)[b]{\Large{\Red{$2$}}}
    \Text(224,192)[b]{\Large{\Red{$3$}}}
    \Text(256,112)[l]{\Large{\Red{$4$}}}
    \Text(224,0)[l]{\Large{\Red{$5$}}}
    \Text(64,0)[l]{\Large{\Red{$6$}}}
    \Text(144,112)[]{\Large{\Red{$7$}}}
    \Text(32,144)[lb]{\Large{\Black{$k1$}}}
    \Text(144,208)[lt]{\Large{\Black{$k2$}}}
    \Text(240,144)[lb]{\Large{\Black{$k3$}}}
    \Text(240,32)[b]{\Large{\Black{$k4$}}}
    \Text(144,-16)[lb]{\Large{\Black{$k5$}}}
    \Text(32,32)[lb]{\Large{\Black{$k6$}}}
    \Text(101,168)[lb]{\Large{\Black{$\rput[lb]{-56}{k2-k1}$}}}
    \Text(159,126)[lb]{\Large{\Black{$\rput[lb]{56}{k3-k2}$}}}
    \Vertex(80,0){3}
    \Vertex(208,0){3}
    \Vertex(256,96){3}
    \Line[arrow,arrowpos=0.5,arrowlength=5,arrowwidth=2,arrowinset=0.2](80,192)(144,96)
    \Line[arrow,arrowpos=0.5,arrowlength=5,arrowwidth=2,arrowinset=0.2](208,192)(144,96)
    \Line[arrow,arrowpos=0.5,arrowlength=5,arrowwidth=2,arrowinset=0.2](256,96)(144,96)
    \Line[arrow,arrowpos=0.5,arrowlength=5,arrowwidth=2,arrowinset=0.2](208,0)(144,96)
    \Line[arrow,arrowpos=0.5,arrowlength=5,arrowwidth=2,arrowinset=0.2](80,0)(144,96)
    \Line[arrow,arrowpos=0.5,arrowlength=5,arrowwidth=2,arrowinset=0.2](32,96)(144,96)
    \Text(64,99)[lb]{\Large{\Black{$k1-k6$}}}
    \Text(176,99)[lb]{\Large{\Black{$k4-k3$}}}
    \Text(96,32)[lb]{\Large{\Black{$\rput[lb]{56}{k6-k5}$}}}
    \Text(165,71)[lb]{\Large{\Black{$\rput[lb]{-56}{k5-k4}$}}}
  \end{picture}
}
\end{center}

\section{Additional features}

\subsection{Integrating at and above the threshold}

Although the sector decomposition strategies are guaranteed to resolve
singularities only if all terms of the function $F$ are of the same sign, {\tt
FIESTA} can sometimes succeed in a wider range of the problems. {\tt FIESTA}
tries to automatically find squares of differences of variables inside $F$ and
to make special integration region decompositions before the sector
decomposition. As a result, the code might succeed in evaluating Feynman
integrals for (a) Feynman integrals at threshold. (b) Feynman integrals for the
three-loop static potential. One of the examples recently calculated is the
following:

\begin{center}
\fcolorbox{white}{white}{
  \begin{picture}(352,198) (3,-11)
    \SetWidth{1.0}
    \SetColor{Black}
    \Line[arrow,arrowpos=0.5,arrowlength=5,arrowwidth=2,arrowinset=0.2](96,86)(128,86)
    \Line[arrow,arrowpos=0.93,arrowlength=5,arrowwidth=2,arrowinset=0.2](128,86)(320,182)
    \Line[arrow,arrowpos=0.93,arrowlength=5,arrowwidth=2,arrowinset=0.2](128,86)(320,-10)
    \Line[dash,dashsize=10](256,150)(256,22)
    \Line[dash,dashsize=10](224,134)(224,38)
    \Line[dash,dashsize=10](288,166)(192,54)
    \Text(320,166)[lb]{\Large{\Black{$q_1^2=M^2$}}}
    \Text(320,-10)[lb]{\Large{\Black{$q_2^2=M^2$}}}
    \Text(0,86)[lb]{\Large{\Black{$(q_1+q_2)^2=4M^2$}}}
  \end{picture}
}
\end{center}

This is a three-loop non-planar vertex at the threshold $(q_1+q_2)^2=4M^2$.
All solid lines are
massive with mass $M$. The integrals of this type are used, for example, in
heavy-fermion corrections to the three-loop matching coefficient of the vector
current\cite{MPSS}.

One might also try to produce the first poles of integrals above the threshold.
To do that one must ensure that the {\tt UsingC} option is set to {\tt False}.
Although one should use those results with care, we do not provide any guarantee
for the correctness in this case.

\subsection{Analytical results}

Another reason to turn {\tt UsingC} to {\tt False} is to evaluate the highest
poles analytically. In this case one should also set the
{\tt ExactIntegrationOrder} to the maximal $\varepsilon$ order that {\tt
Mathematica} should try to evaluate analytically and probably to change the
timeout time {\tt ExactIntegrationTimeout} for a single sector integral. The
default value is $10$. After a timeout {\tt Mathematica} proceeds with
numerical evaluation.

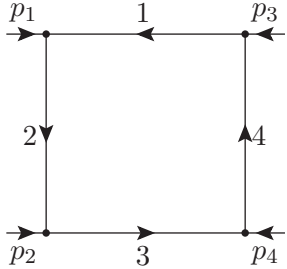
\begin{figure}[ht]
\begin{center}
\fcolorbox{white}{white}{
  \begin{picture}(135,125) (45,-5)
    \SetWidth{0.5}
    \SetColor{Black}
    \ArrowLine(75,95)(75,20)
    \ArrowLine(75,20)(150,20)
    \ArrowLine(150,20)(150,95)
    \ArrowLine(150,95)(75,95)
    \Vertex(75,20){1.41}
    \Vertex(75,95){1.41}
    \Vertex(150,95){1.41}
    \Vertex(150,20){1.41}
    \ArrowLine(165,95)(150,95)
    \ArrowLine(165,20)(150,20)
    \ArrowLine(60,95)(75,95)
    \ArrowLine(60,20)(75,20)
    \Text(150,15)[lt]{\small{\Black{$\;p_4$}}}
    \Text(75,15)[rt]{\small{\Black{$p_2\;$}}}
    \Text(112,15)[ct]{\small{\Black{$3$}}}
    \Text(75,57)[rc]{\small{\Black{$2\;$}}}
    \Text(150,57)[lc]{\small{\Black{$\;4$}}}
    \Text(150,100)[lb]{\small{\Black{$\;p_3$}}}
    \Text(75,100)[rb]{\small{\Black{$p_1\;$}}}
    \Text(112,100)[cb]{\small{\Black{$1$}}}
  \end{picture}
}
\end{center}
\caption{
\label{box}
The massless on-shell box diagram.
}
\end{figure}

For example, for the box diagram (Fig.~\ref{box}) with Mandelstam variables
equal to  $s=-3$ and $t=-1$ and {\tt UsingC=False} the code normally returns
\[
(1.333304 \pm 0.000019)*\varepsilon^{-2}+(-0.732465 \pm 0.000667)*\varepsilon^{-1}+(-4.386581 \pm 0.000767)
\]

If one sets {\tt ExactIntegrationOrder} to {\tt 0}, the result is
\[
(4/3)*\varepsilon^{-2}+((-2*\log[3])/3)*\varepsilon^{-1}+(-4.386649 \pm 0.00018)
\]

Moreover, setting {\tt ExactIntegrationTimeout} to {\tt 60} changes the result
once more and we obtain
\[
(4/3)*\varepsilon^{-2}+((-2*\log[3])/3)*\varepsilon^{-1}+((-4*Pi^2)/9)
\]
Of course, the minimal required timeout depends not only on the complexity of
the problem but also on the CPU speed.

\subsection{Different integrators}

The original {\tt FIESTA} used a {\tt Fortran} implementation of {\tt Vegas} as
the integrator. Currently we have plugged in the Thomas Hahn Cuba library
\cite{Cuba}. By default {\tt FIESTA} uses the {\tt Vegas} integrator, but this
behavior can be easily controlled by the user, see the description of
the option {\tt CurrentIntegrator} in sect.\ref{AlgorithmUsage}.

Cuba implements four algorithms for multidimensional numerical
integration: Vegas, Suave, Divonne, and Cuhre. Cuhre is a
deterministic algorithm, the others use Monte Carlo
methods.

Vegas uses importance sampling for variance reduction. It is the
simplest of the four but since the sector decomposition algorithm
produces pretty good integrands the Vegas usually is the best choice.

Suave combines the advantages of two popular methods: importance
sampling as done by Vegas and subregion sampling.  Divonne works by
stratied sampling, where the partitioning of the integration region is
aided by methods from numerical optimization.

Cuhre is a deterministic algorithm, it uses a cubature rule for
subregion estimation in a globally adaptive subdivision scheme. In low
dimensions it may be used in hope to get the answer with high precision.

It is rather straightforward to add a new integrator to the C-code,
see comments in the file ``{\tt integrators.h}''.

\subsection{More general classes of integrals}

The new version of {\tt FIESTA} can be also applied to more general classes of
integrals:
\begin{eqnarray}
 \int_0^1\ldots \int_0^1\dd x_1\ldots \dd x_{n}\prod_{j=1} \left[P_j\right]^{r_j}\;,
\label{more-general}
\end{eqnarray}
 where $P_j$ are non-negative polynomials of $x_i$ and $r_j$ powers linearly dependent
 on a complex parameter $\ep$.

The syntax is:

\noindent {\tt SDEvaluateDirect[x,{P[1],P[2],\ldots,P[n]},{r[1],r[2],\ldots,r[n]}, order]}

By default, {\tt FIESTA} assumes that singularities in $\ep$ arise only from regions of
small values of integration variables $x[1],\ldots,x[n]$.
However, one might specify the variables for which singularities in $\ep$
are generated also at a vicinity of the value $x[i]=1$.
The syntax is: {\tt BisectionVariables=\{i1,i2,i3,i4,i5\}} where $i1,\ldots$ correspond to
such variables.
To be on the safe side, one might list all the variables but this
can decrease the performance essentially.

This new feature of {\tt FIESTA 2} was successfully applied to parametric integrals contributing to
Wilson loops in \cite{DDS} where it was used to check numerically analytic results.

\section{Code installation}

\label{Inst}
In order to install {\tt FIESTA 2}, the user has to download the installation
package 
\\
http://www-ttp.particle.uni-karlsruhe.de/$\sim$asmirnov/data/FIESTA\_2.0.0.tar.gz,
\\
unpack it and follow the instructions in the file INSTALL.

The {\tt Mathematica} part of {\tt FIESTA} requires almost no installation,
one only needs to copy the {\tt FIESTA\_2.0.0.m} file and edit
the default paths QLinkPath, CIntegratePath and DataPath
in this file, for example:
\begin{itemize}
\item
{\tt QLinkPath="/home/user/QLink/QLink"};
\item
{\tt CIntegratePath="/home/user/FIESTA/CIntegrate"};
\item
{\tt DataPath="/tmp/user/temp"};
\end{itemize}
Here \verb|QLinkPath| is a  path to the executable {\tt QLink} file,
\verb|CIntegratePath| is a  path to the executable {\tt CIntegrate}
file, and \verb|DataPath| is a path to the database directory.
For the Windows system, these paths should look like
\begin{itemize}
\item
{\tt QLinkPath="C:/programs/QLink/QLink.exe"}\footnote{{\tt Mathematica} uses normal slashes for paths both in Unix and Windows.};
\item
{\tt CIntegratePath="C:/programs/FIESTA/CIntegrate.exe"};
\item
{\tt DataPath="D:/temp"};
\end{itemize}
Note that the program will create a big IO traffic to the directory
\verb|DataPath|, therefore, it is better to put this directory on a fast
local disk.

Alternatively, one can specify all these paths manually after loading
the file {\tt FIESTA\_2.0.0.m} into {\tt Mathematica}.

Please note that the code requires {\tt Wolfram Mathematica} 6.0  or 7.0
(recommended) to be installed and will not work correctly under lower versions
of {\tt Mathematica}.

In order to work with nontrivial integrals, the user must install
{\tt QLink} and the C-part of {\tt FIESTA}, the {\tt CIntegrate} program.
The {\tt QLink} \cite{QLink} can be downloaded as a binary file or compiled from the
sources. If the user decides to use pre-compiled {\tt CIntegrate} executable
file, he has to place the file to some location and edit the paths
in the file {\tt FIESTA\_2.0.0.m} as it is described above. If the
user wants to compile the executable file himself he must
have several software packages to be installed on his computer.

First, the Mathematica Developer Kit. It should be installed if the
user has the official {\tt Wolfram Mathematica} installation.

The CIntegrate program depends on the MPFR library \cite{MPFR} and the
Cuba library \cite{Cuba}. After installing MPFR and the Cuba library
the user has to edit the self-explanatory Makefile and run the command
``make''. Then, two executable files should appear, {\tt CIntegrate} and
{\tt CIntegrateMP}. The first one uses the native IEEE floating-point
arithmetic while {\tt CIntegrateMP} uses the MPFR multi-precision
library if necessary. {\tt CIntegrateMP} is slightly slower than
{\tt  CIntegrate} but it should produce a correct result for almost
all integrals, so we strongly recommend to use {\tt CIntegrateMP}

The C-sources are situated in the subdirectory ``sources''. The
program {\tt CIntegrate} is compiled in the subdirectory ``native''
and the program {\tt CIntegrateMP} is compiled in the subdirectory
``mpfr''. After successful compilation both executable files are moved
to the root FIESTA directory. In order to clean up the directory
structure, the user may use the command ``make clean''.

Under Windows the compilation should be performed under
the Cygwin environment. In this case, the executable files will get
the extension ``.exe''.

\section{Algorithm usage}
\label{AlgorithmUsage}

To run {\tt FIESTA} load the {\tt FIESTA\_2.0.0.m} into {\tt Wolfram Mathematica} 6.0 or 7.0
To evaluate a Feynman integral one has to use the command

{\tt SDEvaluate[\{U,F,l\},indices,order]},

where {\tt U} and {\tt F} are the functions defined by (\ref{Dform}) and (\ref{Aform}),
{\tt l} is the number of loops,
{\tt indices} is the set of indices and {\tt order} is the required
order of the $\varepsilon$-expansion.

There is a special syntax that is required to use the Speer sectors strategy, for details see section \ref{SectionSpeer}

To avoid manual construction of $U$ and $F$ one can use a build-in function {\tt UF}
and launch the evaluation as

{\tt SDEvaluate[UF[loop\_momenta,propagators,subst],indices,order]},

where {\tt subst} is a set of substitutions for external momenta, masses and
other values (note, the code performs numerical integrations. Thus the
functions {\tt U} and {\tt F}
should not depend on any external values).

To expand an integral by some variable tending to zero from the positive side, use the command

{\tt SDExpand[\{U,F,l\},indices,order,var,var order]},

where {\tt var} is the expansion variable and {\tt var order} is the required variable expansion order.
One should not care about the possible degrees of logarithms arising, they will be accounted automatically.
Again it is possible to run the code with

{\tt SDExpand[UF[loop\_momenta,propagators,subst],indices,order,var,var order]}

Now the variable replacements should result in $U$ and $F$ depending only on {\tt var}.

Examples:

{\tt SDEvaluate[UF[\{k\},\{-k$^2$,-(k+p$_1$)$^2$,-(k+p$_1$+p$_2$)$^2$,-(k+p$_1$+p$_2$+p$_4$)$^2$\},
\\
\{p$_1^2\rightarrow$0,p$_2^2\rightarrow$0,p$_4^2\rightarrow$0,
p$_1$ p$_2\rightarrow$-s/2,p$_2$ p$_4\rightarrow$-t/2,p$_1$ p$_4\rightarrow$-(s+t)/2,
\\
s$\rightarrow$-3,t$\rightarrow$-1\}],
\{1,1,1,1\},0]
}

performs the evaluation of the massless on-shell box diagram
(Fig.~\ref{box}) where the Mandelstam variables are equal to  $s=-3$ and $t=-1$.

{\tt SDExpand[UF[\{k\},\{-k$^2$,-(k+p$_1$)$^2$,-(k+p$_1$+p$_2$)$^2$,-(k+p$_1$+p$_2$+p$_4$)$^2$\},
\\
\{p$_1^2\rightarrow$0,p$_2^2\rightarrow$0,p$_4^2\rightarrow$0,
p$_1$ p$_2\rightarrow$-s/2,p$_2$ p$_4\rightarrow$-t/2,p$_1$ p$_4\rightarrow$-(s+t)/2,
\\
s$\rightarrow$-1,t$\rightarrow$-tt\}],
\{1,1,1,1\},0,tt,0]
}
expands the corresponding integral.

{\tt SDEvaluateDirect[x, {P[1],P[2],... ,P[n]},
  {r[1],r[2],... ,r[n]}, order]}
evaluates the integral
$\int_0^1\ldots \int_0^1\dd x_1\ldots \dd x_{n}\prod_{j=1} \left[P_j\right]^{r_j}\,.$

To clear the results from memory use the {\tt ClearResults[]} command.

Here are the main options of the code (concerning high-precision
arithmetics options see the end of the sect.~\ref{highPrecision}). One
should set the values before running the {\tt SDEvaluate} command,
e.g. {\tt NumberOfLinks=8}):

\begin{itemize}
\item {\tt UsingC}: specifies whether the C-integration should be used; default value: {\tt True};
\item {\tt CIntegratePath}: path to the {\tt CIntegrate} binary;
\item {\tt NumberOfLinks}: number of the {\tt CIntegrate} programs to be launched; default value: $1$;
\item {\tt NumberOfSubkernels}: number of {\tt Mathematica} subkernels used; default value: $0$ (no {\tt Mathematica} parallelization;
\item {\tt UsingQLink}: specifies whether {\tt QLink} should be used to store data on disk;
works only with {\tt UsingC=True}; default value: {\tt True};
\item {\tt QLinkPath}: path to the {\tt QLink} binary;
we strictly recommend to use the new {\tt QLink} binary working with the TokyoCabinet database and not with QDBM;
\item {\tt DataPath}: path to the place where {\tt QLink} stores the data;
for example, if {\tt DataPath=/temp/temp}, then the
code creates two files: {\tt /temp/temp1} and
{\tt /temp/temp2}; those files will be erased if existent;
the directory {\tt /temp} should exist;
\end{itemize}

Our recommendation for most complicated enough problems is to have
{\tt UsingC} and {\tt UsingQLink} set to {\tt True}; the {\tt  NumberOfLinks} and {\tt NumberOfSubkernels} should be equal to the
number of kernels on your computer; {\tt Mathematica} 7.0 and the new version of {\tt QLink} should be used. Also one should ensure that the {\tt DataPath} directory does not point to a network drive.

There are other options that one can use:

\begin{itemize}
\item {\tt MixSectors}: the number of sectors, that have to be
  integrated together. We do not recommend touching this option with
  {\tt UsingC=True}, but if the {\tt Mathematica} integration is on, one might set it to something big or even {\tt Infinity} to optimize the integration speed;
\item {\tt NegativeTermsHandling}: the way the algorithm should deal with negative terms before the sector decomposition. Currently there are two options: {\tt "Squares"} (default) and {\tt "None"};
\item {\tt CurrentIntegrator}: the integrator used in C. Currently there are four options: {\tt "vegasCuba"} (default option), {\tt "suaveCuba"}, {\tt "divonneCuba"} and {\tt "cuhreCuba"};
\item {\tt CurrentIntegratorSetting}: the options of the current integrator; run the job with default setting and you will see the current settings in the beginning of the output; they can copied to {\tt Mathematica} and edited there. For details on those options see \cite{Cuba}];
\item {\tt PrimarySectorCoefficients}: The usage of this option allows
  to take the symmetries of the diagram into account. If the diagram has symmetries,
then the primary sectors corresponding to symmetrical lines result in equal
contributions to the integration result.
Hence it makes sense to speed up the calculation by specifying the
coefficients before the primary sector integrands. For example, if two lines in the
diagram are symmetrical, one can have a zero coefficient before one of those
and $2$ before the second. {\tt PrimarySectorCoefficients} defines those
coefficients if set; the size of this list should be equal to the number of primary sectors;
\item {\tt STRATEGY}: defines which sector decomposition strategy is used;
{\tt STRATEGY\_0} is not exactly a strategy, but an instruction not to perform the sector decomposition;
{\tt STRATEGY\_A} and {\tt STRATEGY\_B} are the two strategies from Ref.~\cite{BognerWeinzierl} guaranteed to terminate;
{\tt STRATEGY\_S} (default value) is our strategy, producing better results than the preceding ones;
{\tt STRATEGY\_SS} is the strategy based on Speer sectors, one has to provide the diagram structure to use it;
{\tt STRATEGY\_X} is an heuristic strategy from \cite{BognerWeinzierl};
likely to share the ideas of Binoth and Heinrich \cite{BH}: powerful but not guaranteed to terminate;
{\tt ResolutionMode}: the method of dealing with singularities after
the sector decomposition. Possible values: {\tt "IBP0"}, {\tt "IBP1"}
and {\tt "Taylor"} (default value)\footnote{There is no any reason to
  use anything except {\tt "Taylor"}with multiple precision integration
since the algorithm every time will select the ``native''
arithmetics and the only result is an overhead for evaluating of
(\ref{testmonom}) in each point};
\item {\tt RemoteLinksFile}: the file where the code reads the information about remote slave {\tt CIntegrate} jobs that can be used;
\item {\tt RemoteLinksInstallTimeout}: if a remote link cannot be installed after this amount of seconds, the code ignores the link;
\item {\tt RemoteLinksTimeout}: a timeout for reading a ready result from a remote link; after such a timeout the frozen link is banned for a minute for the first time and twice more each next time;
\item {\tt ExactIntegrationOrder}: the maximal order of $\varepsilon$ where {\tt FIESTA} tries to present an analytical result; works only if {\tt UsingC} is set to {\tt False}.
\item {\tt ExactIntegrationTimeout}: the timeout for such an integration attempt for a single sector integral. The default value is $10$. After a timeout {\tt Mathematica} proceeds with numerical evaluation.
\item {\tt d0}: the dimension of the space-time when $\varepsilon\rightarrow 0$; default value is $4$;
\item {\tt ReturnErrorWithBrackets}: if {\tt True}, the code returns the error estimates in the form {\tt pm[57]} and not {\tt pm57};
\item {\tt BisectionVariables}
lists the variables for which the resolution of the singularities in
$\ep$ has to take into account a vicinity of the value $x[i]=1$. It is applied only
with {\tt SDEvaluateDirect}.
\end{itemize}

The following options are left for backward compatibility, but we recommend to avoid using them:
\begin{itemize}
\item {\tt IntegrationCut}: the actual low boundary of the integration domain (instead of zero); default value: $0$;
\item {\tt IfCut} and {\tt VarExpansionDegree}: if {\tt IfCut} is nonzero then the expression
is expanded up to order {\tt VarExpansionDegree} over some of the integration variables; the integration
function is evaluated exactly if the integration variable is greater that {\tt IfCut},
otherwise the expansion is taken instead; the default value of {\tt IfCut} is zero,
the default value of {\tt VarExpansionDegree} is $1$;
\end{itemize}

The following options were used in {\tt FIESTA 1} but have been completely removed from that moment:

\begin{itemize}
\item {\tt ForceMixingOfSectors}: see {\tt MixSectors}
\item {\tt PrepareStringsWhileIntegrating}: no analogue
\item {\tt ResolveNegativeTerms}: see {\tt NegativeTermsHandling}
\item {\tt VegasSettings}: see {\tt CurrentIntegrator} and {\tt CurrentIntegratorSettings}
\end{itemize}

Some remarks on the usage of the code:

\begin{itemize}
\item In complicated cases one should use the {\tt QLink} in order to store expressions
on disk, otherwise there is a good chance to result in memory overflow;
\item {\tt Wolfram Mathematica} is sometimes very greedy in using RAM, to minimize this
usage we minimize the internal {\tt Mathematica} cache and keep erasing in
constantly. If your jobs require this cache for optimization, then it is
advised to run them in a separate {\tt Mathematica} session
\item The package compilation results in two {\tt CIntegrate} binary files ---
with the multiprecision and without it. Although the multiprecision binary is
slower, it is advised to use it to avoid the numerical instability problems
(even if you are aiming at a few digits for your integral, the high precision
arithmetics is required in between, see section \ref{highPrecision}).
\item For complicated integrals it makes sense to increase the number of integral
evaluations performed during the integration (otherwise the algorithm might
fail to adapt to all peaks and underestimate the error). The default value is
$50 000$, while the {\tt FIESTA 1} default value was $1 500 000$. Such
a small number leads to very fast integration but the obtained value of
the integral is usually only a rough estimation which is nevertheless
enough to check numerically the
correctness of the known analytical result. This setting can be
adjusted by the option {\tt CurrentIntegratorSettings}. To get
a real answer with several reliable digits one should set something
like ${\tt CurrentIntegratorSettings}=$

\noindent $\{\{"maxeval","1500000"\},\{"nincrease","10000"\}$.
%\begin{eqnarray}
%\tt CurrentIntegratorSettings=\nonumber &&\\
%\{\{"maxeval","1500000"\},\{"nincrease","10000"\}\}\nonumber
%\end{eqnarray}
\end{itemize}

\section{Numerical results}

In \cite{BaikovCriterion} a set of the
four-loop massless propagator master integrals was identified.
Analytical results in
expansion in $\varepsilon$ for these integrals will be published soon
\cite{BaCh}. Using FIESTA we have calculated\footnote{
A paper is in preparation.} the most complicated integrals of this family and obtained a
good agrement with the analytical results. Here we consider one of the
most complicated diagrams which is shown in Fig.~\ref{m61}. As usually, 
it is implied within {\tt FIESTA} that Feynman integrals are with the $-k^2-i 0$ dependence
of propagators and results are presented, in a Laurent expansion in $\ep$,
by pulling out the factor $i \pi^{d/2} e^{-\gm_E \ep}$ per loop, where $\gm_E$
is the Euler constant.

\begin{table}[!ht]
\begin{minipage}{\linewidth}
\begin{tabular}{|c|c|c|c|}
\hline
Degree & Exact & Cuba Vegas & Cuba Vegas\\
of $\varepsilon$:&Value:&500 000 result:&1 500 000 result:\\
\hline
\hline
      $\varepsilon^{-1}$ & -10.3692776  & -10.36941 $\pm$ 0.00011& -10.36931 $\pm$ 0.00006\\
      $\varepsilon^{0}$ & -70.99081719  & -70.989 $\pm$ 0.002& -70.990 $\pm$ 0.0011\\
      $\varepsilon^{1}$ & -21.663005  & -21.633 $\pm$ 0.023& -21.650 $\pm$ 0.013\\
      $\varepsilon^{2}$ & ---  & 2832.86 $\pm$ 0.17 & 2832.69 $\pm$ 0.096(\footnote{
Calculated with the FORTRAN VEGAS using 1 550 000 samples.})\\
\hline
\end{tabular}
\end{minipage}
\caption{\label{BaikovChetIntsVal} Numerical results for one of the most complicated
four-loop propagator master integrals. In the second column,
the numerical values of the known analytical results are shown. The last two columns
contain the results of evaluation of these integrals by FIESTA using the Cuba VEGAS integrator
with 500000 and 1500000 sampling points. The integral is calculated up to
one extra power of $\varepsilon$  which is unknown analytically.}
\end{table}

\begin{table}[!ht]
\begin{minipage}{\linewidth}
\begin{tabular}{|c|r@{.}l|r@{.}l|}
\hline
Degree &      \multicolumn{2}{|c|}{Cuba Vegas} & \multicolumn{2}{|c|}{Cuba Vegas}\\
of $\varepsilon$:&\multicolumn{2}{|c|}{500 000 time:}&\multicolumn{2}{|c|}{1 500 000 time:}\\
\hline
\hline
  $\varepsilon^{-1}$ & 2262&86s &3495 &28s\\
  $\varepsilon^{0}$  & 15242&10s &39673 &48s\\
  $\varepsilon^{1}$  & 61481&36s &162453 &52s\\
  $\varepsilon^{2}$  & 202018&31s&1794640&00s(\footnote{
Integration by the FORTRAN VEGAS using 1 550 000 samples.})\\
    \cline{1-5}
  Total time: &768727 &00s &\multicolumn{2}{|c|}{---}\\
\hline
\end{tabular}
\end{minipage}
\caption{\label{BaikovChetIntsTime} Timing for calculations of the
  most complicated master integral. The last two columns contain times (in seconds) of numerical
  integration by the Cuba VEGAS integrator with 500000 and 1500000
  sampling points. Also a total time for evaluation of the integral
  is given, including the Mathematica part.}
\end{table}

We use the Cuba VEGAS
integrator with different parameters  for the numerical integration.
Comparing with the analytical results (Table~\ref{BaikovChetIntsVal}), the restriction
in \mbox{500 000} sampling points
leads to the numerical result with 3-4 reliable digits in a quite
reasonable time (see Table~\ref{BaikovChetIntsTime}) while the
integration with \mbox{1 500 000} sampling points reproduces the analytical
results with 4-5 digits. We also have evaluated one extra
$\varepsilon$ term expansion which is unavailable analytically.  For
some technical reasons, for the highest $\varepsilon$ term of the
integral with \mbox{1 500 000} sampling points,
we have restricted ourselves with the value produced
by the FORTRAN VEGAS integrator which is not supported anymore.

\section{Conclusion}

We have presented an essentially updated version of our algorithm.
It can be now used not only for evaluating Feynman integrals by
sector decomposition but also for expanding Feynman integrals in
various limits of momenta and masses.  We have demonstrated the
new features with examples. We believe {\tt FIESTA 2} to be an
important tool for practical calculations and cross-checks of
analytical results.

\vspace{0.2 cm}

{\em Acknowledgments.}
We are grateful to K.G.~Chetyrkin for useful discussions and to 
P.~Marquard and M.~Steinhauser
for testing our code on numerous examples.
This work was supported in part by DFG through SBF/TR 9 and the
Russian Foundation for Basic Research through grant 08-02-01451.


\begin{thebibliography}{99}


\bibitem{Hepp}
K.~Hepp,
%{\em Proof of the Bogolyubov-Parasiuk theorem on renormalization,
Commun. Math. Phys. {\bf 2} (1966) 301.
%%CITATION = CMPHA,2,301;%%

\bibitem{theory}
E.~R.~Speer,
%{\em Analytic renormalization,
J. Math. Phys.,  {\bf 9} (1968) 1404; \\
%\bibitem{Bergere}
M.~C.~Berg\`ere and J.~B.~Zuber,
%{\em Renormalization of Feynman amplitudes and parametric integral
%representation,
Commun. Math. Phys.  {\bf 35} (1974) 113; \\
%%CITATION = CMPHA,35,113;%%
M.~C.~Berg\`ere and Y.~M.~Lam,
%{\em Bogolyubov--Parasiuk theorem in the alpha parametric representation,
J. Math. Phys. {\bf 17} (1976) 1546; \\
%%CITATION = JMAPA,17,1546;%%
O.~I.~Zavialov, {\em Renormalized quantum field theory}, Kluwer Academic Publishers,
Dodrecht (1990);
\\
V.~A.~Smirnov,
%{\em Asymptotic expansions in limits of large momenta and masses,
Commun. Math. Phys. {\bf 134} (1990) 109.
%%CITATION = CMPHA,134,109;%%

\bibitem{BM}
P.~Breitenlohner and D.~Maison,
%{\em Dimensional renormalization and the action principle,
Commun. Math. Phys. {\bf 52} (1977) 11; 39,55; \\
%%CITATION = CMPHA,52,11;%%
%{\em Dimensionally renormalized Green's functions for theories with massless
%particles,
%Commun. Math. Phys.}, {\bf 52} (1977) 39, 55.
%\bibitem{books1}

\bibitem{BdCMPo}
M.~C.~Berg\`ere, C.~de Calan and A.~P.~C.~Malbouisson,
%{\em A theorem on asymptotic expansion of Feynman amplitudes,
Commun. Math. Phys.  {\bf 62} (1978) 137; \\
%%CITATION = CMPHA,62,137;%%
%\bibitem{Po}
K.~Pohlmeyer,
%{\em Large momentum behavior of the Feynman amplitudes in the $\phi^4$ in
%four-dimensions theory,
J. Math. Phys. {\bf 23} (1982) 2511.
%%CITATION = JMAPA,23,2511;%%

\bibitem{books1a}
V.~A.~Smirnov, {\em Applied asymptotic expansions in momenta and masses}, STMP
{\bf 177}, Springer, Berlin, Heidelberg (2002).
%%CITATION = STPHB,177,1;%%

\bibitem{Speer2}
E.~R.~Speer,
%{\em Ultraviolet and infrared singularity structure of generic Feynman
%amplitudes,
Ann. Inst. H. Poincar\'e, {\bf 23} (1977) 1.
%%CITATION = AHPAA,23,1;%%

\bibitem{BH}
T.~Binoth and G.~Heinrich, Nucl. Phys. B, {\bf 585} (2000) 741;
%[hep-ph/0004013];
%T.~Binoth and G.~Heinrich,
Nucl. Phys. B, {\bf 680} (2004) 375;
%[hep-ph/0305234];
%T.~Binoth and G.~Heinrich,
Nucl. Phys. B, {\bf 693} (2004) 134.
%[hep-ph/0402265]

\bibitem{Heinrich}
G.~Heinrich, Int. J. of Modern Phys. A,  {\bf 23} (2008) 10.
[arXiv:0803.4177].

\bibitem{3box}
V.~A.~Smirnov,
%{\em Analytical result for dimensionally regularized massless on-shell planar
%triple box,
Phys. Lett. {\bf B 567} (2003) 193
[arXiv:hep-ph/0305142].
%%CITATION = PHLTA,B567,193;%%

\bibitem{3loop}
T.~Gehrmann, G.~Heinrich, T.~Huber and C.~Studerus,
%{\em Master integrals for massless three-loop form factors: one-loop and
%two-loop insertions,
Phys. Lett.  B {\bf 640} (2006) 252
[arXiv:hep-ph/0607185];
%%CITATION = PHLTA,B640,252;%%
\\
G.~Heinrich, T.~Huber and D.~Maitre,
%{\em Master integrals for fermionic contributions to massless three-loop form
%factors,
Phys. Lett. {\bf B 662} (2008) 344
[arXiv:0711.3590 [hep-ph]].
%%CITATION = PHLTA,B662,344;%%

\bibitem{4loop}
R.~Boughezal and M.~Czakon,
%{\em Single scale tadpoles and O(G(F) m(t)**2 alpha(s)**3) corrections to  the
%rho parameter,
Nucl. Phys. {\bf B 755} (2006) 221
[arXiv:hep-ph/0606232].
%%CITATION = NUPHA,B755,221;%%

\bibitem{BognerWeinzierl}
C.~Bogner and S.~Weinzierl,
%{\em Resolution of singularities for
%multi-loop integrals,
Comput. Phys. Commun.  {\bf 178} (2008) 596
[arXiv:0709.4092 [hep-ph]];
%%CITATION = CPHCB,178,596;%%
%{\em Blowing up Feynman integrals,
Nucl. Phys. Proc. Suppl.  {\bf
183} (2008) 256 [arXiv:0806.4307 [hep-ph]].
%%CITATION = NUPHZ,183,256;%%

\bibitem{FIESTA}
A.~V.~Smirnov and M.~N.~Tentyukov,
%{\em Feynman Integral Evaluation
%by a Sector decomposiTion Approach (FIESTA),
Comput. Phys. Commun. {\bf 180} (2009) 735 [arXiv:0807.4129 [hep-ph]].
%%CITATION = ARXIV:0807.4129;%%

\bibitem{FIESTA-appl}
  A.~V.~Smirnov, V.~A.~Smirnov and M.~Steinhauser,
  %``Fermionic contributions to the three-loop static potential,''
  Phys.\ Lett.\  B {\bf 668}, 293 (2008)
  [arXiv:0809.1927 [hep-ph]]; \\
  %%CITATION = PHLTA,B668,293;%%
%\cite{Bonciani:2008wf}
%\bibitem{Bonciani:2008wf}
  R.~Bonciani and A.~Ferroglia,
  %``Two-Loop QCD Corrections to the Heavy-to-Light Quark Decay,''
  JHEP {\bf 0811}, 065 (2008)
  [arXiv:0809.4687 [hep-ph]]; \\
  %%CITATION = JHEPA,0811,065;%%
%\cite{Kiyo:2008mh}
%\bibitem{Kiyo:2008mh}
  Y.~Kiyo, D.~Seidel and M.~Steinhauser,
  %``${\cal O}(\alpha\alpha_s)$ corrections to the $\gamma t\bar{t}$ vertex at
  %the top quark threshold,''
  JHEP {\bf 0901}, 038 (2009)
  [arXiv:0810.1597 [hep-ph]];\\
  %%CITATION = JHEPA,0901,038;%%
%\cite{Bell:2008ws}
%\bibitem{Bell:2008ws}
  G.~Bell,
  %``NNLO corrections to inclusive semileptonic B decays in the shape-function
  %region,''
  Nucl.\ Phys.\  B {\bf 812}, 264 (2009)
  [arXiv:0810.5695 [hep-ph]];\\
  %%CITATION = NUPHA,B812,264;%%
%\cite{Velizhanin:2008pc}
%\bibitem{Velizhanin:2008pc}
  V.~N.~Velizhanin,
  %``Leading transcedentality contributions to the four-loop universal anomalous
  %dimension in N=4 SYM,''
  arXiv:0811.0607 [hep-th];\\
  %%CITATION = ARXIV:0811.0607;%%
%\cite{Ueda:2009xx}
%\bibitem{Ueda:2009xx}
  T.~Ueda and J.~Fujimoto,
  %``New implementation of the sector decomposition on FORM,''
  arXiv:0902.2656 [hep-ph];\\
  %%CITATION = ARXIV:0902.2656;%%
%\cite{Seidel:2009qp}
%\bibitem{Seidel:2009qp}
  D.~Seidel,
  %``Threshold corrections to the gamma t anti-t vertex at O(alpha alpha(s)),''
  arXiv:0902.3267 [hep-ph];;\\
  %%CITATION = ARXIV:0902.3267;%%
%\cite{Heinrich:2009be}
%\bibitem{Heinrich:2009be}
  G.~Heinrich, T.~Huber, D.~A.~Kosower and V.~A.~Smirnov,
  %``Nine-Propagator Master Integrals for Massless Three-Loop Form Factors,''
  Phys.\ Lett.\  B {\bf 678}, 359 (2009)
  [arXiv:0902.3512 [hep-ph]];\\
  %%CITATION = PHLTA,B678,359;%%
%\cite{Baikov:2009bg}
%\bibitem{Baikov:2009bg}
  P.~A.~Baikov, K.~G.~Chetyrkin, A.~V.~Smirnov, V.~A.~Smirnov and M.~Steinhauser,
  %``Quark and gluon form factors to three loops,''
  Phys.\ Rev.\ Lett.\  {\bf 102}, 212002 (2009)
  [arXiv:0902.3519 [hep-ph]];\\
  %%CITATION = PRLTA,102,212002;%%
%\cite{Gluza:2009mj}
%\bibitem{Gluza:2009mj}
  J.~Gluza, K.~Kajda, T.~Riemann and V.~Yundin,
  %``New results for loop integrals: AMBRE, CSectors, hexagon,''
  PoS A {\bf CAT08}, 124 (2008)
  [arXiv:0902.4830 [hep-ph]];\\
  %%CITATION = POSCI,ACAT08,124;%%
%\cite{Bekavac:2009gz}
%\bibitem{Bekavac:2009gz}
  S.~Bekavac, A.~G.~Grozin, D.~Seidel and V.~A.~Smirnov,
  %``Three-loop on-shell Feynman integrals with two masses,''
  Nucl.\ Phys.\  B {\bf 819}, 183 (2009)
  [arXiv:0903.4760 [hep-ph]];\\
  %%CITATION = NUPHA,B819,183;%%
%\cite{Bonciani:2009nb}
%\bibitem{Bonciani:2009nb}
  R.~Bonciani, A.~Ferroglia, T.~Gehrmann and C.~Studerus,
  %``Two-Loop Planar Corrections to Heavy-Quark Pair Production in the
  %Quark-Antiquark Channel,''
  JHEP {\bf 0908}, 067 (2009)
  [arXiv:0906.3671 [hep-ph]];\\
  %%CITATION = JHEPA,0908,067;%%
%\cite{Czakon:2009zw}
%\bibitem{Czakon:2009zw}
  M.~Czakon, A.~Mitov and G.~Sterman,
  %``Threshold Resummation for Top-Pair Hadroproduction to
  %Next-to-Next-to-Leading Log,''
  Phys.\ Rev.\  D {\bf 80}, 074017 (2009)
  [arXiv:0907.1790 [hep-ph]];\\
  %%CITATION = PHRVA,D80,074017;%%
%\cite{Ferroglia:2009ep}
%\bibitem{Ferroglia:2009ep}
  A.~Ferroglia, M.~Neubert, B.~D.~Pecjak and L.~L.~Yang,
  %``Two-loop divergences of scattering amplitudes with massive partons,''
  arXiv:0907.4791 [hep-ph];
  %%CITATION = ARXIV:0907.4791;%%
%\cite{Ferroglia:2009ii}
%\bibitem{Ferroglia:2009ii}
%  A.~Ferroglia, M.~Neubert, B.~D.~Pecjak and L.~L.~Yang,
  %``Two-loop divergences of massive scattering amplitudes in non-abelian gauge
  %theories,''
  JHEP {\bf 0911}, 062 (2009)
  [arXiv:0908.3676 [hep-ph]];\\
  %%CITATION = JHEPA,0911,062;%%
%\cite{Bekavac:2009zc}
%\bibitem{Bekavac:2009zc}
  S.~Bekavac, A.~G.~Grozin, P.~Marquard, J.~H.~Piclum, D.~Seidel and M.~Steinhauser,
  %``Matching QCD and HQET heavy-light currents at three loops,''
  arXiv:0911.3356 [hep-ph];\\
  %%CITATION = ARXIV:0911.3356;%%
%\cite{Dowling:2009md}
%\bibitem{Dowling:2009md}
  M.~Dowling, J.~Mondejar, J.~H.~Piclum and A.~Czarnecki,
  %``Radiative-nonrecoil corrections of order alpha^2 (Z alpha)^5 to the Lamb
  %shift,''
  arXiv:0911.4078 [hep-ph];\\
  %%CITATION = ARXIV:0911.4078;%%
%\bibitem{SSS}
  A.~V.~Smirnov, V.~A.~Smirnov and M.~Steinhauser,
  %``Three-loop static potential,''
 [arXiv:0911.4742 [hep-ph]].

\bibitem{BS}
M.~Beneke and V.~A.~Smirnov,
%``Asymptotic expansion of Feynman integrals near threshold,''
Nucl.\ Phys.\ B {\bf 522}, 321 (1998); \\
%[hep-ph/9711391].
%%CITATION = HEP-PH 9711391;%%
V.~A.~Smirnov and E.~R.~Rakhmetov,
%``The regional strategy in the asymptotic expansion of two-loop vertex  Feynman
%diagrams,''
Theor.\ Math.\ Phys.\  {\bf 120}, 870 (1999) [Teor.\ Mat.\ Fiz.\
{\bf 120}, 64 (1999)]; \\
%[hep-ph/9812529];
%%CITATION = HEP-PH 9812529;%%
V.~A.~Smirnov,
%``Problems of the strategy of regions,''
Phys.\ Lett.\ B {\bf 465}, 226 (1999).
%[hep-ph/9907471].
%%CITATION = HEP-PH 9907471;%%

\bibitem{Pilipp}
V.~Pilipp,
%``Semi-numerical power expansion of Feynman integrals,''
JHEP {\bf 0809} (2008) 135.
%[arXiv:0808.2555 [hep-ph]].
%%CITATION = JHEPA,0809,135;%%

\bibitem{MB}
V.~A.~Smirnov, Phys. Lett.  B  {\bf 460}  (1999) 397;\\
%%CITATION = HEP-PH 9905323;%%
J.~B.~Tausk, Phys. Lett.  B  {\bf 469}  (1999) 225;\\
%%CITATION = HEP-PH 9909506;%%
M.~Czakon, Comput.\ Phys.\ Commun.\  {\bf 175} (2006) 559; \\
%%CITATION = CPHCB,175,559;%%
A.~V.~Smirnov and V.~A.~Smirnov,
%``On the Resolution of Singularities of Multiple Mellin-Barnes Integrals,''
arXiv:0901.0386 [hep-ph].
%%CITATION = ARXIV:0901.0386;%%

\bibitem{books2}
V.~A.~Smirnov, {\em Evaluating Feynman Integrals}, Springer Tracts
Mod.\ Phys.\  {\bf 211} (2004) 1; \\
%%CITATION = STPHB,211,1;%%
V.~A.~Smirnov, {\em Feynman integral calculus},
%\href{/spires/find/hep/www?irn=6927351}{SPIRES entry}
Berlin, Germany: Springer (2006) 283~p.

\bibitem{DP}
%\bibitem{Roth:1996pd}
M.~Roth and A.~Denner, {\it High-energy approximation of one-loop {F}eynman
  integrals},  {\em Nucl. Phys.} {\bf B479} (1996) 495--514,
  [{\tt hep-ph/9605420}];\\
%\bibitem{Denner:2003wi}
A.~Denner, M.~Melles and S.~Pozzorini,
%``Two-loop electroweak angular-dependent logarithms at high energies,''
Nucl.\ Phys.\ B {\bf 662} (2003) 299
[hep-ph/0301241];\\
%%CITATION = HEP-PH 0301241;%%
%\bibitem{Pozzorini:2004rm}
S.~Pozzorini,
%``Next-to-leading mass singularities in two-loop electroweak singlet form
%factors,''
Nucl.\ Phys.\ B {\bf 692}, 135 (2004)
[hep-ph/0401087];\\
%%CITATION = HEP-PH 0401087;%%
A.~Denner and S.~Pozzorini,
Nucl.\ Phys.\  B {\bf 717} (2005) 48
[arXiv:hep-ph/0408068];\\
%%CITATION = NUPHA,B717,48;%%
%\bibitem{Denner:2006jr}
A.~Denner, B.~Jantzen and S.~Pozzorini,
%``Two-loop electroweak next-to-leading logarithmic corrections to  massless
%fermionic processes,''
Nucl.\ Phys.\  B {\bf 761} (2007) 1
[arXiv:hep-ph/0608326];
%%CITATION = NUPHA,B761,1;%%
%\bibitem{Denner:2008hw}
%A.~Denner, B.~Jantzen and S.~Pozzorini,
%``Two-loop electroweak Sudakov logarithms for massive fermion scattering,''
arXiv:0801.2647 [hep-ph];
%%CITATION = ARXIV:0801.2647;%%
%\bibitem{Denner:2008yn}
%A.~Denner, B.~Jantzen and S.~Pozzorini,
%``Two-loop electroweak next-to-leading logarithms for processes involving
%heavy quarks,''
JHEP {\bf 0811} (2008) 062
[arXiv:0809.0800 [hep-ph]];
%%CITATION = JHEPA,0811,062;%%

\bibitem{SDforRR}
%\bibitem{Anastasiou:2003gr}
  C.~Anastasiou, K.~Melnikov and F.~Petriello,
  %``A new method for real radiation at NNLO,''
  Phys.\ Rev.\  D {\bf 69} (2004) 076010
  [arXiv:hep-ph/0311311];
  %%CITATION = PHRVA,D69,076010;%%
%\bibitem{Anastasiou:2004qd}
%  C.~Anastasiou, K.~Melnikov and F.~Petriello,
  %``Real radiation at NNLO: e+ e- --> 2jets through O(alpha(s)**2),''
  Phys.\ Rev.\ Lett.\  {\bf 93} (2004) 032002
  [arXiv:hep-ph/0402280];
  %%CITATION = PRLTA,93,032002;%%
%\bibitem{Anastasiou:2004xq}
%  C.~Anastasiou, K.~Melnikov and F.~Petriello,
  %``Higgs boson production at hadron colliders: Differential cross sections
  %through next-to-next-to-leading order,''
  Phys.\ Rev.\ Lett.\  {\bf 93} (2004) 262002
  [arXiv:hep-ph/0409088];
  %%CITATION = PRLTA,93,262002;%%
%\bibitem{Anastasiou:2005qj}
%  C.~Anastasiou, K.~Melnikov and F.~Petriello,
  %``Fully differential Higgs boson production and the di-photon signal through
  %next-to-next-to-leading order,''
  Nucl.\ Phys.\  B {\bf 724} (2005) 197
  [arXiv:hep-ph/0501130];
  %%CITATION = NUPHA,B724,197;%%
%\bibitem{Anastasiou:2005pn}
%  C.~Anastasiou, K.~Melnikov and F.~Petriello,
  %``The electron energy spectrum in muon decay through O(alpha**2),''
  JHEP {\bf 0709} (2007) 014
  [arXiv:hep-ph/0505069];\\
  %%CITATION = JHEPA,0709,014;%%
%\bibitem{Heinrich:2006ku}
  G.~Heinrich,
  %``The sector decomposition approach to real radiation at NNLO,''
  Nucl.\ Phys.\ Proc.\ Suppl.\  {\bf 157} (2006) 43
  [arXiv:hep-ph/0601232];
  %%CITATION = NUPHZ,157,43;%%
%\bibitem{Heinrich:2006sw}
%  G.~Heinrich,
  %``Towards e+ e- --> 3jets at NNLO by sector decomposition,''
  Eur.\ Phys.\ J.\  C {\bf 48} (2006) 25
  [arXiv:hep-ph/0601062].
  %%CITATION = EPHJA,C48,25;%%

\bibitem{Kaneko:2009qx}
  T.~Kaneko and T.~Ueda,
  %``A geometric method of sector decomposition,''
  arXiv:0908.2897 [hep-ph].
  %%CITATION = ARXIV:0908.2897;%%

\bibitem{Horoi}
M.~Horoi and R.~J.~Enbody,
International Journal of High
Performance Computing Applications, {\bf 15}, No.~1, 75--80 (2001).

\bibitem{Amdahl}
G.~M.~Amdahl, {\em Validity of the Single Processor Approach to
  Achieving Large-Scale Computing Capabilitie},
In: Proc. AFIPS Conf., 1967, Reston, VA, USA.

\bibitem{SS09}
A.~V.~Smirnov and V.~A.~Smirnov,
JHEP, {\bf 05} (2009) 004
%``Hepp and Speer Sectors within Modern Strategies of Sector Decomposition,''
[arXiv:0812.4700 [hep-ph]].
%%CITATION = ARXIV:0812.4700;%%

\bibitem{BaikovCriterion}
  P.~A.~Baikov,
  %``A practical criterion of irreducibility of multi-loop Feynman  integrals,''
  Phys.\ Lett.\  B {\bf 634}, 325 (2006)
  [arXiv:hep-ph/0507053].
  %%CITATION = PHLTA,B634,325;%%

\bibitem{BaCh}
P.~A.~Baikov and K.~G.~Chetyrkin, in preparation.

\bibitem{dimreg1}
G.~'t Hooft and M.~Veltman, Nucl.~Phys. {\bf B44} (1972) 189.
%%CITATION = NUPHA,B44,189;%%

\bibitem{dimreg2}
C.~G.~Bollini and J.~J.~Giambiagi, Nuovo Cim. {\bf 12B} (1972) 20.
%%CITATION = NUCIA,B12,20;%%

\bibitem{Cuba}
T.~Hahn,
%Cuba?a library for multidimensional numerical integration
Comput. Phys. Commun. 168 (2005) 78
[arXiv: hep-ph/0404043].

\bibitem{QLink}
QLink -- open-source program by A.V.~Smirnov,
http://qlink08.sourceforge.net

\bibitem{MPFR}
GNU MPFR  http://www.mpfr.org/ is a portable C library for
arbitrary-precision binary floating-point computation with correct
rounding, based on GMP library http://gmplib.org/

\bibitem{MPSS}
 P.~Marquard, J.~H.~Piclum, D.~Seidel and M.~Steinhauser,
  %``Completely automated computation of the heavy-fermion corrections to the
  %three-loop matching coefficient of the vector current,''
  Phys.\ Lett.\  B {\bf 678}, 269 (2009)
  [arXiv:0904.0920 [hep-ph]].
  %%CITATION = PHLTA,B678,269;%%

\bibitem{DDS}
  V.~Del Duca, C.~Duhr and V.~A.~Smirnov,
  %``An Analytic Result for the Two-Loop Hexagon Wilson Loop in $\N = 4$ SYM,''
  arXiv:0911.5332 [hep-ph].
\end{thebibliography}
\end{document}